\documentclass[draftclsnofoot,12pt,onecolumn] {IEEEtran}

\usepackage{mdwlist}

\usepackage{flushend}
\usepackage{url}
\usepackage{graphicx}
\usepackage{subfigure}
\usepackage{xcolor,colortbl}
\usepackage[mathscr]{eucal}
\usepackage{amsthm }
\usepackage{amssymb}

\usepackage{cases}
\newtheorem{heuristic}{Heuristic}
\newtheorem{definition}{Definition}
\newtheorem{lemma}{Lemma}

\usepackage{mathptmx}

\ifCLASSOPTIONcompsoc
\else
\fi
\ifCLASSINFOpdf
\else
\fi
\begin{document}

%
\title{\huge A PRQ Search Method for Probabilistic  Objects}
%
%
%
%

\author{Jack~Wang 
\IEEEcompsocitemizethanks{\IEEEcompsocthanksitem Contact information: cszjwang@gmail.com. 
\protect\\

}
\thanks{}}

\IEEEcompsoctitleabstractindextext{%
\begin{abstract}
 This article proposes an PQR search  method for probabilistic objects.  
The main idea of our method is to use a strategy called \textit{pre-approximation} that can reduce  the initial problem to a highly simplified version, implying that it  makes the rest of steps easy to tackle. In particular, this strategy itself is pretty simple and easy to implement. Furthermore, motivated by  the cost analysis,   we  further optimize our solution. The optimizations are mainly based on two insights: (\romannumeral 1)  the number of \textit{effective subdivision}s is no more than 1; and (\romannumeral 2) an entity with the larger \textit{span} is more likely to subdivide a single region. We demonstrate the effectiveness and efficiency of our proposed approaches through extensive experiments under various experimental settings. 
\end{abstract}

}

\maketitle

\IEEEdisplaynotcompsoctitleabstractindextext

%
\IEEEpeerreviewmaketitle




\section{Introduction}\label{sec:1}
\begin{figure}[t]
  \centering
  \subfigure[\scriptsize {} ]{\label{fig:6:c}
     \includegraphics[scale=.4]{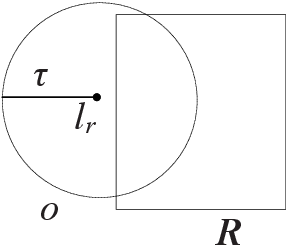}} 
  \subfigure[\scriptsize {} ]{\label{fig:6:a}
     \includegraphics[scale=.4]{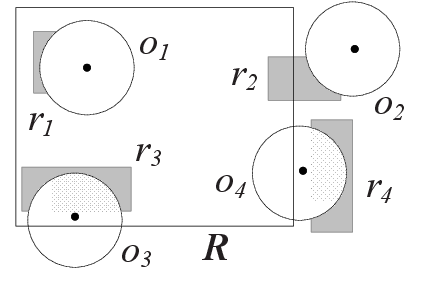}} 
  \subfigure[\scriptsize {}]{\label{fig:6:b}
      \includegraphics[scale=.41]{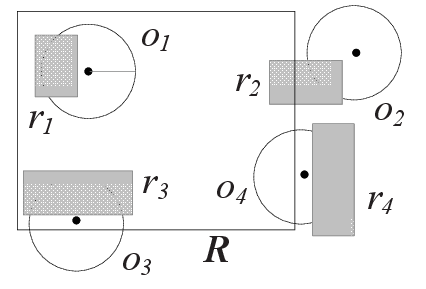}} 
 \caption{\small Illustration of the PRQ and CSPRQ. The small black dot   denotes the recorded location $l_r$, the radius of circle denotes the distance threshold $\tau$, the biggest rectangle  denotes {query range} $R$, the small rectangle   denotes the restricted area $r$. } 
 \label{fig:6}
\end{figure}

\PARstart{R}ange query for moving objects   has  been the subject of much attentions \cite{bugragedik:processing,HaiboHu:AGeneric,HodaMokhtar:On,HaojunWang:Processing,KunLungWu:Incremental}, as it can find   applications in various domains such  as the digital battlefield,  mobile workforce management, and transportation industry. 
It is usual that for a moving object $o$, only the discrete location information is stored on the database server,  due to various reasons such as the limited  battery power of mobile devices and  the limited network bandwidth \cite{reynoldcheng:querying}. 
The recorded location  of  $o$ can be obtained by accessing the database,  the whereabouts of its current location  is usually uncertain \cite{OuriWolfson:Updating}. For example, a common location update policy called \textit{dead reckoning} \cite{reynoldcheng:querying, OuriWolfson:Updating} is to update the recorded location ${l_r}$  when the deviation between
${l_r}$ and the actual location  of  $o$ is larger
than a given distance threshold $\tau$. Before the next update, the specific location of  $o$ is uncertain, except knowing that it lies in a circle  with the center $l_r$ and radius $\tau$.
To capture the location uncertainty, the idea of incorporating \textit{uncertainty} into moving objects data has been proposed \cite{OuriWolfson:Updating}.  From then on, the probabilistic range query (PRQ) as a variant of  the traditional \textit{range query} has attracted much attentions in the data management community    \cite{brucesechung:processing,aprasadsistal:Modeling,dieterpfoser:capturing,jinchuanchen:efficient,gocetrajcevski:managing,reynoldcheng:querying,meihuizhang:effectively, gocetrajcevski:Uncertain}. A well known  \textit{uncertainty model}   is using a closed region (in which the object can always be found) together with a {probability density function} (PDF). The closed region is usually called  \textit{uncertainty region}, and the PDF is used to denote object's \textit{location  distribution} \cite{reynoldcheng:querying,jinchuanchen:efficient,OuriWolfson:Updating}.   (See Section \ref{sec:3} for  a more formal definition.)  Given a query range $R$, the main difference between  the traditional range query and the PRQ is that the latter returns not only the objects being  located in $R$ but also their appearance probabilities.   Assume that  the location of object $o$  follows uniform distribution in its uncertainty region $u$ for ease of discussion, the  probability of object $o$ being located in $R$ is  equal to the ratio of the two areas, i.e., the probability  $p=\frac{area~of~ u\cap R}{area~of~u}$.  Figure \ref{fig:6:c} illustrates an example and the PRQ  returns \{ ($o$, 39\%) \}. 

In existing works, an important branch is to address the PRQ for objects moving freely in 2D space. In this branch, many \textit{uncertainty models}  and  techniques are proposed for various purposes. (Section \ref{sec:2} gives a brief survey about those models, purposes and techniques.)  
Surprisingly, little efforts are made for the PRQ over  objects moving {in a constrained 2D space} where  objects are forbidden to be located in some \textit{specific areas}. For clarity, we term such specific areas as   \textit{restricted areas}, and dub the query above the Constrained Space Probabilistic Range Query (CSPRQ).  The CSPRQ can also find many applications as objects moving in a constrained 2D space are common in the real world. For example, the tanks in the digital battlefield  usually cannot run in  lakes, forests and the like, the areas occupied by those obstacles  can be naturally regarded as \textit{restricted areas} (of tanks). With similar observations, in a zoo, tourists usually cannot roam in the dwelling spaces of dangerous animals such as tigers and lions, those dwelling spaces   can be regarded as the \textit{restricted areas}  (of tourists).

Existing solutions cannot be directly applied to the CSPRQ as it involves a  set ${\mathscr{R}}$  of \textit{restricted areas}. Imagine if we directly use  existing methods, implying that we ignore each restricted area $r$ ($\in {\mathscr R}$) in the  computation phase.   Figure \ref{fig:6:a} depicts this case,   the circle $o.\odot$ is regarded as the uncertainty region $u$, and the query answer is \{($o_1$, 100\%), ($o_3$, 56\%), ($o_4$, 42\%)\}.  In contrast, Figure \ref{fig:6:b} presents the case considering  ${\mathscr R}$ in the computation phase,  here  $o.\odot-\bigcup _{r\in R} r$  is regarded as $u$, and the query answer is \{ ($o_1$, 100\%), ($o_3$, 22\%), ($o_4$, 76\%)\}. 
The  two answers above are different, and clearly the second one is   correct.  At first sight, to process the CSPRQ is simple as it seems to be a straightforward adaptation of existing methods.  The fact however is not so, as  this idea will be confronted with the overcomplicated geometrical operations, rendering  its implementation infeasible.       (Section \ref{sec:frameworkandanalysis} gives  more detailed explanations.) In addition, computing the uncertainty region $u$ is also not a simple subtraction operation, as a straightforward computation  incurs possible  mistakes. On the other hand, the CSPRQ needs to consider a new set ${\mathscr {R}}$ compared to the previous works, it implies that  the amount of data to be processed is larger and the computation is more complicated, which is another challenge and thus needs more considerations.

%


Motivated by the  fact above,  this paper makes the  effort to the CSPRQ. 
The key idea of our solution is to use a strategy called \textit{pre-approximation} that can reduce  the initial problem to a highly simplified version, implying that it  makes the rest of steps easy to tackle. In particular, this strategy itself is pretty simple and easy to implement. To  operate different entities in a unified and efficient manner, a \textit{label based data structure}  is  developed. Ascribing the pre-approximation and label based data structure,  it is pretty simple to compute the appearance probability. To improve the I/O efficiency, a twin index is naturally adopted. Furthermore, motivated by  the cost analysis,   we  further optimize our solution. The optimizations are mainly based on two insights:  (\romannumeral 1)  the number of \textit{effective subdivision}s is no more than 1,  we  utilize this insight to improve the power pruning restricted areas; and  (\romannumeral 2) an entity with the larger \textit{span} is more likely to subdivide a single region, this insight motivates us to sort the entities to be processed according to their {spans}.  In addition to the main insights above, we also realize  two other (simple but usually easy to ignore) facts and utilize them. Specifically, two mechanisms are developed: \textit{postpone processing} and \textit{lazy update}. 
After we finish the main tasks of this work,  we also attempt another approach inspired by the curiosity, its basic idea is to precompute  uncertainty regions  and index them. Unfortunately, this approach suffers a non-trivial preprocessing time although it  outperforms the  aforementioned approaches in terms of both query and I/O performance. This extra finding offers us an important indication sign  for the future research.   
In summary, we make the following contributions.

\begin{itemize*}
\item We  formulate the CSPRQ  based on an \textit{extended} uncertainty model, and analyse its unique properties. 
\item We   show   a straightforward solution will be confronted with non-trivial troubles, rendering its implementation infeasible. We also show   it is  (almost) infeasible to develop an  exact solution.
\item We propose our solution that utilize an (extremely) important but pretty simple strategy.  
\item We further optimize our solution based on two insights and two  (simple but usually easy to ignore) facts. 
\item We demonstrate the performance of our solution through extensive experiments under various experimental settings.
\item We report an extra finding that offers an important indication sign for the future research.
\end{itemize*}







In   the next section we formulate the problem to be studied and review the related work. We  analyse this problem   and propose our  solution in  Section \ref{sec:frameworkandanalysis} and  \ref{sec:our solution}, respectively.  We further optimize our solution in Section \ref{sec:query evaluation}. We attempt the precomputation based approach in Section \ref{sec:indexing}.  We evaluate the efficiency and effectiveness  of our proposed algorithms through extensive experiments in Section \ref{sec:6}.  Finally, Section  \ref{sec:7} concludes this paper with several interesting research topics.


\section{Problem definition} \label{sec:3}
Given a territory $\mathbb{T}$ with a set ${\mathscr R}$ of    disjoint restricted areas, we assume there exist  a set $\mathscr O$ of    moving objects that can freely move in $\mathbb{T}$ but cannot be located in any  restricted area $r$ ($\in {\mathscr R}$), and assume  the last sampled location  of each moving object $o$ is already stored on the database server. (Note that in this paper  the terms the \textit{last sampled location} and  \textit{recorded location} are  used interchangeably.) Moreover, suppose  each  object $o$ reports its new location to the sever when the deviation between the recorded location $l_r$ and the actual  location of $o$  is larger than a given distance threshold $\tau$. 
We denote  the location of $o$ at an arbitrary instant of time $t$ by $l_t$. Furthermore,  for any two different moving objects $o$ and $o^{\prime}$, we assume they cannot be located in the same location at the same instant of time $t$, i.e., $l_t\neq l_t^{\prime}$. Since  the realistic application environment  varies from place to place, the shapes of restricted areas  should be diversified, whereas our objective is  to establish a general approach instead of  focusing on certain specific environment. Therefore throughout this paper we  use polygons to denote the restricted areas (note: this assumption is feasible, since  any shaped  area can be transformed into polygon shaped area beforehand).  Finally, we set the following  conditions are always satisfied: 

\begin{subnumcases}{}
l_t\notin \bigcup_ {r\in {\mathscr R}}{r}   \\
l_t\in {\mathbb{T}}-\bigcup_ {r\in {\mathscr R}}{r} \\
\bigcup_ {r\in {\mathscr R}}{r} \subset {\mathbb{T}}
\end{subnumcases}


The specific location of $o$ at the current time is usually uncertain, a well known model \cite{reynoldcheng:querying,OuriWolfson:Updating} allows us to capture the location uncertainty of $o$ through two components:
\begin{definition}[Uncertainty region]\label{definition:ur}
The {uncertainty region}   of a moving object $o$ at a given time $t$, denoted by $u^t$,  is a closed region where $o$ can always be found.
\end{definition}
\begin{definition}[Uncertainty probability density function]
The {uncertainty probability density function} of a moving object $o$ at a given time $t$, denoted by $f^t(x,y)$, is the  PDF of $o$'s location at the time $t$. Its value is $0$ if $l_t$ $\notin$ $u^t$.
\end{definition}
Note that under the distance based update policy (a.k.a. dead-reckoning
policy \cite{reynoldcheng:querying}), for any two different time $t_1$ and $t_2$,  we have (\romannumeral 1) $ u^{t_1}=u^{t_2}$ and (\romannumeral 2) $ f^{t_1}(x,y)=f^{t_2}(x,y)$, 
where $t_1$, $t_2\in$ ($t_r$, $t_n$],  $t_r$ refers to the latest reporting time, and  $t_n$ refers to the current time. 
In view of these, in the remainder of the paper we  use $u$ and $f(x,y)$ to denote the uncertainty region and  PDF of $o$, respectively.
Since $f(x,y)$ is a PDF, in theory, it has the property:  
\begin{equation}
\int_{{ u}} f(x,y)dxdy=1
\end{equation}
Under the {distance based update} policy, the uncertainty region $u$    can be derived based on the following  formula \cite{reynoldcheng:querying,OuriWolfson:Updating}.
\begin{equation}
u= {{C}} (l_r, \tau)
\end{equation}
where ${{C}}(\cdot)$ denotes a circle with the centre  $l_r$ and  radius 
$\tau$. For convenience, we use $o.\odot$ to denote this region.
The above representation is feasible under the case no restricted areas exist, i.e., ${\mathscr R}=\emptyset$. Whereas the real uncertainty region $u$ for our problem should be as follows.
\begin{equation}
\label{equ:computeURreal}
u=o.\odot-\bigcup _{r\in {\mathscr R}} r
\end{equation}
\begin{definition}[Constrained space probabilistic range query]
Given  a set $\mathscr R$ of restricted areas and a set $\mathscr O$ of moving objects in a territory $\mathbb{T}$, and a  query range $R$, the {constrained space probabilistic range query}  returns a set $\mathscr O^\prime$ ($\subseteq \mathscr O$) of  objects together with their appear probabilities in form of ($o$, $p$) such that for any $o\in \mathscr O^\prime$, $p\neq 0$, where $p$ is the  probability of  ${o}$ being located in  ${R}$, and is computed as $p=\int_{u\cap R} f(x,y)dxdy$.
\end{definition}
Note that  in this paper we assume the distance based update policy is adopted. We abuse the notation '$|\cdot|$' but its meaning should be clear from the context. In addition, a notation or symbol with the subscript 'b' usually refers to its corresponding MBR (e.g., $o.\odot_b$ refers to the MBR of $o.\odot$).  
For ease of reading, we summarize the frequently used symbols in Table \ref{tab:main symbols}.

\begin{table}[h]
\begin{center}
\begin{tabular}{|p{.06\textwidth} p{.36\textwidth}  | }\hline 
{{\footnotesize Symbols}}&{{\footnotesize Description}} \\ \hline 

{\footnotesize $R$}  & {\footnotesize query range} 	\\

{\footnotesize $o$}  & {\footnotesize  moving object}  \\

{\footnotesize $\mathscr O$}  & {\footnotesize the set of moving objects} 	\\

 {\footnotesize $r$}  & {\footnotesize  restricted area}	\\
 
  {\footnotesize $\zeta$}  & {\footnotesize  the number of edges of $r$}	\\
 
  {\footnotesize $\mathscr R$}  & {\footnotesize  the set of restricted areas}	\\

{\footnotesize $\tau$}  & {\footnotesize distance threshold } 	\\

 {\footnotesize $l_r$}  & {\footnotesize the  recorded location of $o$}	\\

  {\footnotesize $f(x,y)$}  & {\footnotesize PDF of  $o$'s location}	\\

{\footnotesize $p$}  & {\footnotesize probability of $o$ being located in  $R$} 	\\

{\footnotesize $u$}  & {\footnotesize uncertainty region}\\

 {\footnotesize $s$}  & {\footnotesize the intersection result between $R$ and $u$}	\\

{\footnotesize $u_o$}  & {\footnotesize the outer ring of $u$} \\

{\footnotesize $\varphi$}  & {\footnotesize the intersection result between $R$ and  $u_o$} 	\\

 {\footnotesize $u_h$}  & {\footnotesize  the hole of $u$}	\\

{\footnotesize $\mathscr H$}  & {\footnotesize the set of holes in $u$}	\\

{\footnotesize $\mathscr O^*$}  & {\footnotesize the set of candidate moving objects} 	\\

{\footnotesize $\mathscr R^*$}  & {\footnotesize  the set of candidate restricted areas}	\\

 {\footnotesize $e$}  & {\footnotesize the approximated equilateral polygon  from $o.\odot$} 	\\

{\footnotesize $\xi$}  & {\footnotesize the number of edges of $e$} 	\\

{\footnotesize $d$}  & {\footnotesize a subdivision}   \\

 {\footnotesize $d^e$}  & {\footnotesize  the effective subdivision}	\\

 \hline
\end{tabular}
\end{center}
\caption{\small symbols and their meanings}\label{tab:main symbols}
\end{table}


\subsection{Related work} \label{sec:2}
In terms of probabilistic range query over uncertain moving objects, researchers have made considerable efforts, and many outstanding techniques and models have been proposed. In this subsection, we review those works most related to ours.  

The uncertainty model used in this paper is developed based on  \cite{OuriWolfson:Updating,reynoldcheng:querying}. In their papers,  a moving object $o$  updates its recorded location $l_r$, when the deviation (between its actual location and $l_r$) is larger than a given distance threshold $\tau$. This update policy  is just the so-called \textit{distance based update policy}{\small \footnote{\small We also assume this update policy is adopted in our work. Another common location update policy is the {time based update}, i.e. updating the recorded location $l_r$ periodically (e.g., every 3 minutes). The   CSPRQ is more challenging if the time based update policy is assumed to be adopted,  as it needs   more considerations on the time dimension and usually needs other assumptions (e.g., the velocity of object should be available). In addition, the space dimension should be  more difficult to handle, as the uncertainty region $u$ is to be a continuously  changing geometry over time. We leave this interesting  topic as the future work, and we believe this paper will lay a foundation for the future research.}}. 
In particular, they discussed two types of moving objects: (\romannumeral 1)   moving on  predefined routes, and (\romannumeral 2)  moving freely in 2D space. For the former, the route consists of a series of line segments,  the uncertainty is a line segment on the route, called line segment uncertainty (LSU) model for convenience. For the latter, the route is unneeded, and  the uncertainty used in their paper is a circle, called free moving uncertainty (FMU) model. Our model  roughly follows the latter. The difference is that our model introduces the  restricted areas, and the uncertainty region $u$ is not necessarily a circle.  (Although only a slight difference viewed from the surface,   the amount of  data to be processed in our query however is  larger, and the computation is more  complicated. In particular, a straightforward adaptation of their method will incur overcomplicated geometrical operations, rendering its implementation infeasible.) 

In addition, the models in \cite{yufeitao:range,jinchuanchen:efficient} are the same or similar as the FMU model, and  also focus on the case of no restricted areas.   For example, 
Tao et al. \cite{yufeitao:range} investigated range query on multidimensional uncertain data, they proposed a classical technique  PCR,  and  an elegant indexing mechanism  U-tree. They  adopted a circle to represent the uncertainty region $u$ (see Section 7 in \cite{yufeitao:range}).  Chen et al. \cite{jinchuanchen:efficient} addressed \textit{location based} range query. Several clever ideas such as query expansion and query duality were proposed. They discussed two types of \textit{target} objects:   static and moving. They  assume the uncertainty region $u$ is a rectangle when the \textit{target} object is moving.  (Note: our work does not belong to \textit{location based}  query.  Location based CSPRQ  should be more interesting, as the location of query issuer is also uncertain.) 

Regarding to the case of  objects   moving freely   in 2D space,  there are many other classical uncertainty models like, the MOST model \cite{aprasadsistal:Modeling}, the UMO model \cite{meihuizhang:effectively}, the 3D cylindrical (3DC) model  \cite{gocetrajcevski:managing,dieterpfoser:capturing}, and the necklace uncertainty (NU) model \cite{gocetrajcevski:Uncertain,BartKuijpers:Trajectory}. These models   have different assumptions and purposes,  but also  their own advantages   (note: it is a difficult task to say which one is the best).  The models in \cite{aprasadsistal:Modeling,meihuizhang:effectively} are developed for  querying  the future location. 
For example, Sistla et al.  \cite{aprasadsistal:Modeling} proposed the MOST model, they assume  the direction and speed of each object $o$ are available, and these information should be updated if the change occurs. The  future location is predicted  based on three parameters: velocity, direction, and time. Later, Zhang et al. \cite{meihuizhang:effectively} proposed the UMO model, in which they use the distribution of location and  the one of velocity, instead of  the exact values, to characterize the location uncertainty, and assume these distributions  are available at the update time.  The models in \cite{gocetrajcevski:managing,gocetrajcevski:Uncertain,dieterpfoser:capturing,BartKuijpers:Trajectory} are suitable for querying the trajectories of moving objects. For example, Trajcevski et al. \cite{gocetrajcevski:managing} proposed to model an uncertain trajectory as a 3D  cylindrical body, they assume an electrical map, all  recorded locations and sampling time are available.  Later, they proposed the NU model  \cite{gocetrajcevski:Uncertain}, which can be viewed as an enhanced version of the 3DC model. In this model, they represent the whereabouts in-between two known locations as a bead, and an uncertain trajectory  as a necklace  (a sequence of beads).   Our work is different from aforementioned works in at least two points:  (\romannumeral 1) those works focus on the case of no restricted areas, and (\romannumeral 2) the underlying uncertainty  model is different from theirs. (Note: it should be more interesting  to extend the concept of restricted areas to those uncertainty models.)

Recently, Emrich et al. \cite{TobiasEmrich:querying} proposed to model the trajectories of  moving objects by stochastic processes, they assume the object is in a \textit{discrete state space} (i.e., a finite set of possible locations in space), and assume the transition probability (from a state to another state) is available. Our work is different form theirs in two points at least: (\romannumeral 1)  the underlying models are different, and (\romannumeral 2)  the object discussed in our paper is not in a discrete state space. 

Another important branch is to focus on  objects  moving on  predefined routes (or road networks) \cite{brucesechung:processing,KaiZheng:probabilistic}. For example, 
Chung et al. \cite{brucesechung:processing}  adopted the LSU model  to process range query, and proposed a clever idea --- transforming the uncertain movements of objects into points in a dual space using the Hough Transform.  To query  the trajectories of objects moving on  road networks, Zheng et al. \cite{KaiZheng:probabilistic} proposed  the uncertain trajectory (UT) model and an elegant indexing mechanism UTH. They assume all  recorded locations and sampling time are available, and   objects  follow the shortest paths and travel at a constant speed between two consecutive trajectory samples. Our work  is different from  works mentioned above, as here we  focus on objects moving in a constrained 2D space where no predefined routes are given.

\section{Problem analysis}\label{sec:frameworkandanalysis}
At first sight, to process the CSPRQ is simple as it seems to be a straightforward adaptation of  existing methods. To process the PRQ, existing methods (see e.g., \cite{reynoldcheng:querying})  consist of three main steps.    
\begin{enumerate*}
\item  For each object $o$, it computes  $u\cap R$ if the uncertainty region $u$ intersects with the query range $R$.
\item   It computes the probability $p$ based on a formula $p=\int _{u\cap R} f(x,y)dxdy$, and put the tuple ($o$, $p$) into the result. 
\item  It returns the result (which usually includes a series of tuples)  after all objects are processed.
\end{enumerate*} 
By the large, we only need to add \textit{one} step, i.e., computing the uncertainty region $u$ based on Equation  \ref{equ:computeURreal} before checking if $u$ intersects with $R$. In other words, this straightforward method  mainly consists of four steps.    Now the readers should be pretty curious --- why the four steps  above cannot be (easily) achieved. 
We next look a bit deeper into those steps above, and then we can easily realize four main issues (but not
limited to) arise. 


\begin{figure} [h]
\centering
  \subfigure[\scriptsize { } ]{\label{fig:12a}
     \includegraphics[scale=.37]{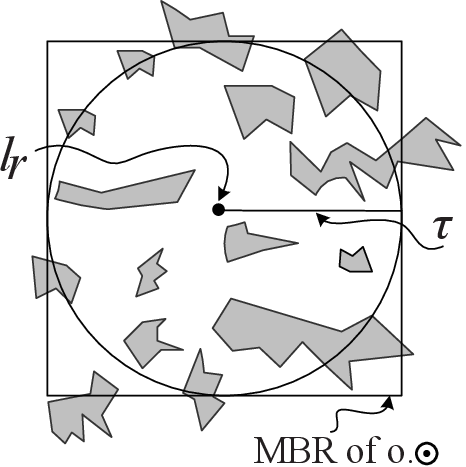}} 
  \subfigure[\scriptsize { }]{\label{fig:12b}
      \includegraphics[scale=.37]{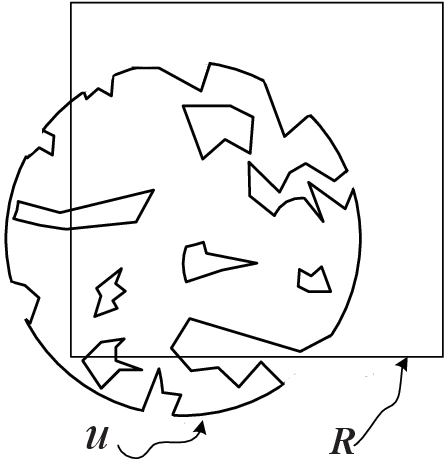}} 
  \subfigure[\scriptsize { }]{\label{fig:12c}
      \includegraphics[scale=.37]{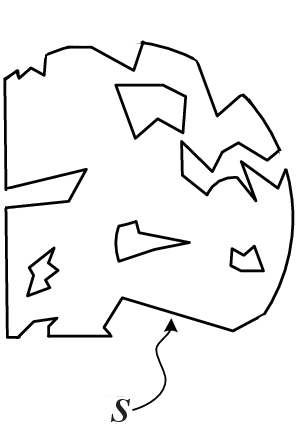}} 
\caption{\small  Illustration of a straightforward  solution. (a) The grey polygon illustrates the restricted area $r$,  the circle illustrates  $C(l_r,~\tau)$, i.e.,  $o.\odot$. (b) The biggest rectangle illustrates the query range $R$, the pseudo circle illustrates the uncertainty region $u$. (c) It is the intersection result of $u\cap R$.} 
 \label{fig:12}
\end{figure}

\begin{figure} [h]
\centering
  \subfigure[\scriptsize { } ]{\label{fig:13a}
     \includegraphics[scale=.37]{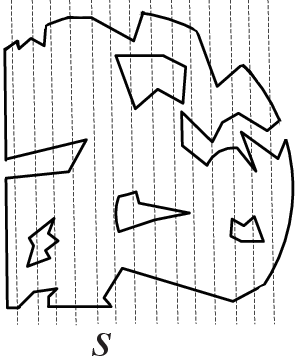}} 
  \subfigure[\scriptsize { }]{\label{fig:13b}
      \includegraphics[scale=.37]{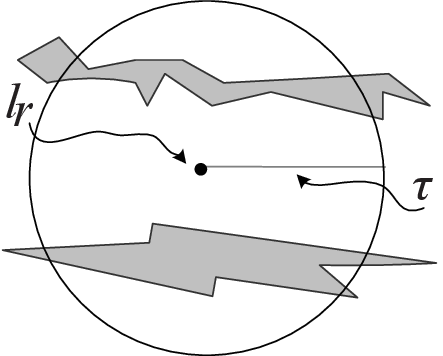}} 
  \subfigure[\scriptsize { }]{\label{fig:13c}
      \includegraphics[scale=.37]{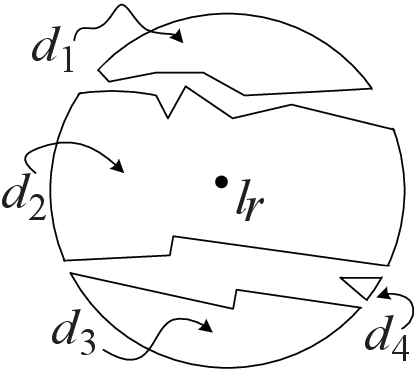}} 
\caption{\small  Illustration of computing the area and  uncertainty region $u$. The grey polygon illustrates the restricted area $r$,  the circle illustrates  $C(l_r,~\tau)$, i.e.,  $o.\odot$.} 
 \label{fig:13}
\end{figure}

First,  suppose the location of an object $o$ follows  {uniform distribution} in its uncertainty region $u$,  the following equation holds \cite{dieterpfoser:capturing}:
\begin{equation}\label{equation:area ratio uniform}
p=\frac{\Lambda(u\cap R)}{\Lambda(u)}
\end{equation}
where $\Lambda(\cdot)$ denotes the  area of the geometrical entity.  Let  $s$ be the \textit{intersection result} of $u\cap R$. It is easy to know that computing the area of $u$ (or $s$) is simple for the case of no restricted areas. To the case of our concern,  e.g., see Figure \ref{fig:12}, how to compute the area of $u$ (or $s$)? 
Computing the area of such complicated  entity is not an easy task, as its boundary consists of both straight line segments and curves, and it includes many holes. (In fact,  $s$ possibly consists of multiple subdivisions  in addition to holes. Those even more complicated cases will be discussed in Section \ref{sec:query evaluation}.)   
A natural method could be to  divide the  entity into multiple small strips shown in Figure \ref{fig:13a}, and then to compute the area of each strip and add them together.  In practice, this solution however,  is  overcomplicated and difficult to implement. 

Second, suppose the location of  $o$ does not follow  uniform distribution in $u$, 
a usually used method   is the  Monte Carlo method. Its basic idea is to randomly generate $N_1$ points in $u$. For each generated point $p^{\prime}$, it computes  $f(x_i,y_i)$, where ($x_i,~y_i$) is the coordinates of the point $p^{\prime}$, and then checks whether or not $p^{\prime}\in s$.  Without loss of generality, suppose  $N_2$ points (among $N_1$ points) are to be located in $s$. Finally, it gets the probability $p$ as follows.
\begin{equation}\label{equ:montecarlo}
p=\frac{\sum_{i=1}^{N_2}{f{(x_i,~y_i)}}}{\sum_{i=1}^{N_1}{f{(x_i,~y_i)}}}
\end{equation}
Given a randomly generated point $p^{\prime}$,   to check whether or not  $p^{\prime}\in u$ (or $p^{\prime}\in s$) is  simple if no restricted areas exist.  However,  it is not an easy task for the case of our concern. Note that the solutions to the \textit{point in polygon problem}    \cite{KaiHormann:ThePointInPolygonProblem,MarkdeBerg:computational} cannot be applied to our context as geometrical entities considered here are more complicated. 


Third, it is easy to know that both $u$ and $s$ are pretty simple if  no restricted areas exist.  Geometrical entities in our context however  are  more  complicated. Then, how to  represent and operate them in a concise and efficient way?  

Fourth,  computing  $u$ is not a straightforward {subtraction operation}.  Figure \ref{fig:13b} illustrates the case before executing the subtraction operation, and the subtraction result is shown in  Figure \ref{fig:13c}, which has four subdivisions. Note that, only  $d_2$ is the real uncertainty region, other subdivisions are invalid, the reason  will be explained in the next section. 

Besides the  issues mentioned above, we should note that the amount of data to be processed is  larger compared to the case of no restricted areas.  It is easy to know that the PRQ only needs to check $O(|\mathscr O|)$ objects. In contrast, the CSPRQ needs to check  $O(|\mathscr R|)$ restricted areas for each object $o$, the (worst case) complexity is $O(|\mathscr R||\mathscr O|)$.








\noindent \textbf{Discussion.} The above analysis offers insights into our problem, it  reveals to us that to process the CSPRQ using a  straightforward solution is infeasible. Furthermore, even if the location of $o$ follows uniform distribution in $u$, it is still non-trivial  to develop an  exact solution (let alone the non-uniform distribution case),  implying that to develop an exact solution  is also (almost) infeasible. 
After we realize the facts above, we also note that another easily brought to mind method that is to approximate the curves on the boundary of $u$ (or $s$) into line segments. In this way, the troubles shown before seemingly can  be   tackled easily. In fact, existing \textit{curve interpolation techniques} can indeed transform the boundary of $u$ (or $s$) into line segments. It however is  still inconvenient and inefficient, since there are too many such entities in the query processing. In addition, it is also difficult and troublesome to approximate curves into line segments in such a manner, as the shapes of  different entities  vary from one to another.

\section{Our solution} \label{sec:our solution}

The  key idea of our solution is to use a strategy called \textit{pre-approximation}   that can lead  to a highly simplified version of the initial problem, implying that it can make  the rest of steps easy to tackle. In particular, this strategy itself is pretty simple and  easy to implement. 


\subsection{Pre-approximation}
{The essence of the pre-approximation is that,   it  first transforms (or approximates) $o.\odot$ into an equilateral polygon denoted by $e$, and then uses $e$ to subtract the restricted areas. Thus, according to Equation \ref{equ:computeURreal}, we have 
\begin{equation}\label{equation:approximationUR}
u\doteq e-\bigcup _{r\in \mathscr R} r 
\end{equation}



To get $e$ is pretty simple, without loss of generality, assume that we need to approximate $o.\odot$ into  an  equilateral polygon $e$ with a number $\xi$ of edges.  We actually only need to obtain each vertex of $e$, which can be computed based on the following equations.   
\begin{subnumcases}{}\label{equ:approximate}
{ x}_i=l_r.{x}+ \tau\cdot\cos \frac{(i-1)\cdot {2\pi}}{\xi}  \\
{ y}_i=l_r.{y}+{\tau}\cdot\sin \frac{(i-1)\cdot {2\pi}}{\xi}\label{equ:approximate1}
\end{subnumcases}
where  $i$$\in$$[1,2,\dots,{\xi}]$,  ($l_r.x$, $l_r.y$) denote the coordinates of the recorded location $l_r$, and $({x}_i,{y}_i)$ denote the coordinates of the $i${th} vertex of $e$.

Clearly, the larger (the) $\xi$ is, the more accurate results we can get. (In fact, our experimental results show the accuracy is pretty good even if we only set $\xi=32$.)  Note that, here  ${o}.\odot$ is the circumscribed circle of  $e$, which can assure that the distance from any point in $e$ to the center  is always less than the distance threshold $\tau$. 
The main reasons we do  this transformation are as follows:   
(\romannumeral 1)  it is convenient for the follow-up calculations since operating on line segments, in most cases, is more simple and efficient than on curves;
(\romannumeral 2) it is easy to represent the calculated result; and (\romannumeral 3)  all the troubles discussed in Section \ref{sec:frameworkandanalysis} can be significantly simplified.  In the next section, we show how to represent different entities in a unified and efficient manner.

\subsection{LBDS}\label{subsec: lbds}


%


Once  the   pre-approximation idea is adopted, the boundaries of all the geometrical entities will be no curves.  Moreover, we observe that $u$ may be a closed region with hole(s) or  just be  a simple closed region, and $s$ possibly consists of multiple subdivisions with hole(s).  For ease of operating these entities in a unified and efficient  manner, we need a targeted data structure to represent   them. (We remark that the  doubly connected edge list (DCEL) \cite{MarkdeBerg:computational}  consists of three collections of records: one for the vertices, one for the faces, and one for the half-edges; to our problem, it is a little clunky and not intuitive enough.) We next introduce some basic definitions  in order to easily describe the details of our proposal.
\begin{definition}[Outer ring and inner ring]\label{definition:hole}
Given a closed region $c$ with a hole $h$ ,  the  boundary of $c$ and the one of $h$ are termed as the outer ring and  inner ring of $c$, respectively. 
\end{definition}
Note that,  if a closed region contains $n$ holes,  it clearly has  $n$ inner rings, see e.g., Figure \ref{fig:12c}. 
Specifically, in order to easily handle those entities, we present a label based data structure (LBDS), which consists of three domains --- one \textit{label} domain and two \textit{pointer} domains. 
\begin{itemize*}
\item \texttt{Flag}: This domain  is the boolean type. Specifically, $0$  indicates  the entity has no hole, and 1 indicates it has no less than one hole. 
\item \texttt{OPointer}: This domain points to  a simple polygon that denotes the outer ring of the entity. A simple polygon consists of two domains.
\begin{itemize*}
\item \texttt{VPointer}: This domain points to a linked list that stores a series of vertexes.
\item \texttt{B}: This domain stores the MBR of the polygon.
\end{itemize*}
\item \texttt{IPointer}: This domain points to a linked list that stores the simple polygons, which denote the inner rings   of this entity.
\end{itemize*}

Hence $u$ can  be  represented  by the LBDS directly, and $s$ can be represented by a linked list in which a series of 'LBDSs' are stored. This structure   is intuitive, concise, and  convenient  for the follow-up computation, its benefits will  be demonstrated  gradually in the rest of the paper. 


\subsection{Picking out the real uncertainty region} \label{subsec:pickout}

To compute the uncertainty region $u$  (by Formula \ref{equation:approximationUR}) is straightforward, we can use the equilateral  polygon $e$ to subtract  each restricted area $r$ one by one. 
In Section \ref{sec:frameworkandanalysis}, we show that computing $u$  is not a simple subtraction operation. In other words,  Formula \ref{equ:computeURreal} and  \ref{equation:approximationUR} actually imply some possible mistakes,  we slightly abuse them for presentation simplicity.  In  Figure \ref{fig:13c}, we say  $d_2$ (rather than other three subdivisions)  is the real uncertainty region, which  is based on the  lemma below.

\begin{lemma}[Choose real uncertainty region]\label{lemma:choose real ur}
Given  $o.\odot$, $l_r$, $\tau$ and $\mathscr R$, we let $d$ be one of subdivisions  after we execute the subtraction operation based on Equation \ref{equ:computeURreal}. If  $l_r \in d$, then $\surd$($d$), where $\surd$($\cdot$) denotes  it is the real uncertainty region. Otherwise, $\lnot$($\surd$($d$)).
\end{lemma}
\noindent \textbf{Proof.}
We first prove $l_r\notin d$ $\Rightarrow$ $\lnot$($\surd$($d$)). According to Definition \ref{definition:ur}, we  only need to show   $o$ cannot be found  in $d$. Clearly, $o$  must be located in  $o.\odot$, since $l_r$  is the latest recorded location, and  $\tau$ is the distance threshold. Furthermore,  based on  \textit{analytic geometry},  it is easy to know that  $o$  cannot reach  $d$  if it does not walk out of   $o.\odot$ (e.g., see Figure \ref{fig:13b}, $o$ cannot reach the topmost (or bottommost) region of $o.\odot$). Hence we cannot find  $o$ in $d$. 

The proof for the argument ``$l_r\in d$ $\Rightarrow$ $\surd$($d$)'' can be obtained using the similar method above; omitted due to space limit.
  $\square$


We remark that once the  pre-approximation idea is used, to check whether or not $l_r\in d$ is simple, as it is just the   \textit{point in polygon problem} \cite{KaiHormann:ThePointInPolygonProblem, MarkdeBerg:computational}. 
After we obtain the real uncertainty region $u$,  we can get $s$ by executing   an  \textit{intersection operation} on $u$ and $R$.  There are  many algorithms (e.g., see \cite{BalaRVatti:aGeneric,GuntherGreiner:efficient,AriRappoport:Anefficient,YongKuiLiu:anAlgorithm,AvrahamMargalit:AnAlgorithm}) that can  perform intersection operation on polygons with holes. They however do not well consider the case of many holes. Even so, there is a simple method that is adapted from the  algorithms mentioned
above. Its general idea is to compute the intersection result between $R$ and the outer ring of $u$ at first, and then to use this intersection result to subtract each inner ring of $u$ one by one, finally it gets $s$.  



\subsection{The appearance probability }\label{subsec:two solvers}


For uniform distribution PDF,  the crucial task is to compute the areas of $u$  and $s$ (cf. Equation \ref{equation:area ratio uniform}). We show in Section \ref{sec:frameworkandanalysis} that computing these areas using a straightforward solution  is  overcomplicated.  Now, we  can easily  compute them using the following method, which ascribes the  pre-approximation and  LBDS.  
Let  $u_o$ be the outer ring  of $u$,  and $u_h^i$ be  the $i$th hole in $u$. Let $s[i]$ be  the $i$th subdivision of $s$, $s[i]_o$ be the outer ring of $s[i]$, and {$s[i]_h^k$} be the  $k$th  hole of {$s[i]$}. 
First, given a polygon denoted by $P$, its area can be easily obtained  based on the following equation  \cite{website:areapolygon}.

\begin{equation}\label{equ:polygonarea}
\Lambda (P)=\frac{1}{2}\cdot \left( 
                        \left|
                           \begin{array}{cc}
                           x_1   &x_2   \\
                           y_1   &y_2
                           \end{array}
                        \right| +                        
                        \left|
                              \begin{array}{cc}
                               x_2   &x_3   \\
                               y_2   &y_3
                               \end{array}
                           \right| + ... +
                           \left|
                           \begin{array}{cc}
                           x_{n}   &x_{1}   \\
                           y_{n}   &y_{1}
                           \end{array}
                        \right|
                     \right) 
\end{equation}
where {$\left|
                           \begin{array}{cc}
                           x_1   &x_2   \\
                           y_1   &y_2
                           \end{array}
                        \right|$}$=(x_1\cdot y_2 -x_2\cdot y_1)$, and $(x_1,y_1)$ 
                        denote the coordinates of a vertex, other symbols have similar meanings. 
Furthermore, since we use the LBDS to represent $u$, and  polygons are the basic elements of the LBDS,  the area of $u$ can be  obtained as follows.
\begin{equation}\label{equ:uarea}
\Lambda(u)=\Lambda(u_o) -\sum_{i=0}^{|\mathscr H|}  \Lambda(u_h^i)
\end{equation}
where  $|\mathscr H|$ ($\geq 0$) is the number of   holes in $u$. Similarly, since $s$ consists of a series of LBDSs, we have
\begin{equation}\label{equ:sarea}
\Lambda(s)= \sum_{i=1}^{|s|}\Lambda(s[i]) =\sum_{i=1}^{|s|} {(\Lambda(s[i]_o) -\sum_{k=0}^{|s[i]_h|}\Lambda(s[i]_h^k)} 
\end{equation} 
where $|s|$ ($\geq 1$) is the number of subdivisions of {$s$},  $|s[i]_h|$ ($\geq 0$) is the number of holes in {$s[i]$}.
For arbitrary distribution PDF, we also use the Monte Carlo method to compute the probability $p$. We should note that the trouble  shown  in Section \ref{sec:frameworkandanalysis} does not exist now, as  no curve is on the boundary of $u$ (or  $s$), ascribing the   pre-approximation idea. 
\subsection{Query processing}\label{subsec:query processing}
A naive method is to do a linear scan ---  for each object $o$, it scans each restricted area $r$, and compute $u$ based on Formula \ref{equation:approximationUR}, and then compute the probability $p$ if $u$ intersects with  $R$.  Clearly, it is  inefficient to process the CSPRQ in such a way. We now present another natural but more efficient method as follows.  Let $R_b$, $r_b$ and $o.\odot_b$  be the MBRs of $R$, $r$ and $o.\odot$, respectively. 
\begin{definition}[Candidate moving object]
Given    the query range $R$ and a moving object $o$, $o$ is a candidate moving object  such that $R_b\cap$$o.\odot_b$$\neq$$\emptyset$.
\end{definition}

\begin{definition}[Candidate restricted area]
Given   a restricted area $r$ and a moving object $o$, $r$ is a candidate restricted area  such that $r_b\cap$$o.\odot_b$$\neq$$\emptyset$.
\end{definition}
Let $\mathscr O^*$ be the set of candidate moving objects, and $\mathscr R^*$ be the set of candidate restricted areas of the object $o$. We can rewrite Formula \ref{equation:approximationUR} as follows.
\begin{equation}
\label{equation:modifiedURCRA}
u\doteq e-\bigcup _{r\in \mathscr R^*} r 
\end{equation}
The MBRs of the set $\mathscr R$ of restricted areas can be obtained easily, since each restricted area $r$ is static. Furthermore, since the recorded location $l_r$ and the distance threshold $\tau$ 
are already stored on the database server, the MBR of each moving object can  be 
 computed easily, it is  a square centering at $l_r$ and   with  2$\tau$ $\times$ 2$\tau$ size (in fact it is just the MBR of  $o.\odot$). Clearly, for all restricted areas and moving objects, we can 
use a twin-index  to manage their MBRs. For instance, we can build two R-trees (or a variant such as the R$^*$- tree) to manage the MBRs of  moving objects and the ones of restricted areas, respectively. Let $\mathscr I_o$ and $\mathscr I_r$ be the index of moving objects and the one of restricted areas, respectively.  
Our query processing algorithm is illustrated below.

\vspace{1ex}

{ \hrule

\small

\noindent \textbf{Algorithm 1} {\footnotesize Constrained space probabilistic range query} 
\hrule

}
{ 



    {\footnotesize

(1)$~~~~$Let  $\Re=\emptyset$

(2)$~~~~$Search  $\mathscr O^*$ on $\mathscr I_o$ using $R_b$ as the input 

(3)$~~~~$\textbf{for} each $o\in \mathscr O^*$ \textbf{do}

(4)$~~~~$$~~~~$Search  $\mathscr R^*$ on $\mathscr I_r$ using $o.\odot_b$ as the input 
    
(5)$~~~~$$~~~~$Obtain  $e$ based on Equation \ref{equ:approximate} and \ref{equ:approximate1}  

(6)$~~~~$$~~~~$Compute $u$ based on Formula \ref{equation:modifiedURCRA}    

(7)$~~~~$$~~~~$\textbf{if} $u$ consists of multiple subdivisions \textbf{then}

(8)$~~~~$$~~~~$$~~~~$Choose the real $u$ based on Lemma \ref{lemma:choose real ur} 

(9)$~~~~$$~~~~$Let $s=u\cap R$, and $p=0$

(10)$~~~$$~~~~$\textbf{if} $s\neq \emptyset$ \textbf{then}

(11)$~~~$$~~~~$$~~~~$\textbf{if} uniform distribution PDF  \textbf{then} 

(12)$~~~$$~~~~$$~~~~$$~~~~$Compute $p$ based on Equation \ref{equation:area ratio uniform}, \ref{equ:polygonarea}, \ref{equ:uarea} and \ref{equ:sarea}

(13)$~~~$$~~~~$$~~~~$\textbf{else} // non-uniform distribution PDF 

(14)$~~~$$~~~~$$~~~~$$~~~~$Compute $p$ based on Equation \ref{equ:montecarlo}

(15)$~~~$$~~~~$\textbf{if} $p\neq 0$ \textbf{then}

(16)$~~~$$~~~~$$~~~~$Let $\Re=\Re\bigcup(o,p)$

(17)$~~~$\textbf{return}  $\Re$
\hrule  
        }

\vspace{1ex}
\noindent \textbf{Cost analysis.} Let $C_o$  be the  cost to search the set $\mathscr O^*$ of candidate moving objects. Clearly, we have 
\begin{equation}\label{formu:costo}
C_o\propto  (|R_b|,~|\mathscr O|)
\end{equation}
where $\propto$ means ``is proportional to'', $|R_b|$ is the size of MBR of $R$, $|\mathscr O|$ is the cardinality of $\mathscr O$. Let $C_r$ be the  cost to search  the set $\mathscr R^*$ of candidate restricted areas. Similarly, we have  
\begin{subnumcases}{}
C_r\propto (|o.\odot_b|,~|\mathscr R|) \label{formu:costr1}    \\
|o.\odot_b|\propto \tau  \label{formu:costr2}
\end{subnumcases}
where $|o.\odot_b|$ is the size of MBR of $o.\odot$, $|\mathscr R|$ is the cardinality of $\mathscr R$, $\tau$ is the distance threshold of $o$. Let $C_e$ be the cost to obtain the equilateral polygon $e$. We have 
\begin{equation}
C_e\propto \xi
\end{equation}
where $\xi$ is the number of edges of $e$.  
Let $C_u$ be the  cost to compute  $u$ (line 6-8), and $\zeta$ be  the average number of edges of  restricted areas. We have
\begin{subnumcases}{}
C_u\propto (\xi,~\zeta,~|\mathscr R^*|)    \\
|\mathscr R^*| \propto (\tau,~|\mathscr R|) 
\end{subnumcases}
where $|\mathscr R^*|$ is the cardinality of $\mathscr R^*$. Let $C_s$ be the  cost to compute $s$, and $\gamma$ be the number of edges of $u_o$ (the outer ring of $u$). Since  the set $\mathscr R^*$ of candidate restricted areas  form the holes of $u$, the  number of edges of hole is also $\zeta$. We have 
\begin{subnumcases}{}
C_s\propto (\gamma,~|u_h|,~\zeta)    \\
|u_h|\propto (\tau,~|\mathscr R^*|) 
\end{subnumcases}

Let $C_p^u$ be the  cost to compute $p$ in the case of uniform distribution, and $C_p^n$ be the  cost to compute $p$ in the case of non-uniform distribution.  Note that  each cost (in the {for} loop) refers to the average cost, and we overlook the cost of adding a tuple $(o,p)$ into $\Re$ as it is trivial. We also note that $C_o$ is related to $\mathscr I_o$ (e.g., the fan-out of $\mathscr I_o$), and $C_r$ is related to $\mathscr I_r$ (e.g., the fan-out of $\mathscr I_r$).  We assume  existing indexing technique is to be adopted, we hence omit this discussion in our analysis for simplicity.  Let $C_t$ denote the total cost, we have
\begin{subnumcases}{C_t=}\label{equ:cost1}
C_o+ |\mathscr O^*|({C_r}+C_e+C_u+C_s+C_p^u) \\
C_o+ |\mathscr O^*|({C_r}+C_e+C_u+C_s+C_p^n)    \label{equ:cost2}
\end{subnumcases}
Clearly, to reduce the total cost $C_t$, we should reduce at least one  sub-cost. Since $|\mathscr O|$ and $|\mathscr R|$ are depended on the application scenario, and $|R_b|$ is depended on the input of the user, we can easily know by Formula \ref{formu:costo}, \ref{formu:costr1} and \ref{formu:costr2} that there is little space to reduce $C_o$ and $C_r$. Furthermore, there is also (almost) no space to reduce $C_e$, as $\xi$ is used to assure the accuracy of our algorithm, and the natural solution to execute Equation \ref{equ:approximate} and \ref{equ:approximate1} is already pretty efficient. We also note that our solution to compute the area of $u$ (or $s$) is already pretty simple and efficient, implying that there is also (almost) no  space to reduce $C_p^u$. Regarding to $C_p^n$, it is mainly depended on the number $N_1$ of random generated points (cf. Equation \ref{equ:montecarlo}),  and $N_1$ is used to assure the accuracy of our algorithm. Naturally, we  need to set $N_1$ to an acceptable value at least which can assure an allowable workload error. This implies that there is also no much space to reduce $C_p^n$. 
Recall Section \ref{subsec:pickout}, we compute $u$ and $s$ using the simple methods, which are somewhat inefficient, and  for each single query, the cost to compute $u$ and $s$ is $|\mathscr O^*|(C_u+C_s)$, which is non-trivial compared to $C_t$. (Note that in the previous works,  $C_u=0$,  $C_s$ is pretty   small and almost can be overlooked, as $u$ is a circle in the case of no restricted areas.)   These facts motivate us to further optimize our solution by reducing $C_u$ and $C_s$. 
In the next section, we  show how to reduce $C_u$ and $C_s$ based on two insights and two simple  but usually  easy to ignore facts. 

\section{Further optimize our solution} \label{sec:query evaluation}



The optimizations are mainly based on two insights: (\romannumeral 1)  the number of \textit{effective subdivision}s is no more than 1; and (\romannumeral 2) an entity with the larger \textit{span} is more likely to subdivide a single region. In addition to the main insights above, we also realize  two other (simple but usually easy to ignore)  facts and utilize them;  specifically, two mechanisms are developed: \textit{postpone processing} and \textit{lazy update}. 

\subsection{Effective subdivision} \label{subsec:effectiveSubdivision}

\begin{definition}[Effective subdivision]\label{definition:effective subdivision}
Given  $o.\odot$ and a set $\mathscr R^*$ of candidate restricted areas,  without loss of generality,  assume that $|{o.\odot}-r|>1$ when  the $i$\textup{th} ``subtraction operation'' is executed, where  $|\cdot|$ denotes the number of subdivisions of the subtraction result, and $1\leq i\leq |\mathscr R^*|$.  A  subdivision $d$ is an effective subdivision such that the recorded location $l_r\in d$.  
\end{definition}

\begin{lemma}[Number of effective subdivisions]\label{lemma:numbersubdivision1}
Assume that $|o.\odot-r|>1$, the number of effect subdivisions is no more than 1.
\end{lemma}
\noindent \textbf{Proof.}
It follows from Definition \ref{definition:effective subdivision} and \textit{analyst geometric}, the details are omitted due to space limit.
$\square$


Let $d^e$ be the effective subdivision when  the $i$th ``subtraction operation'' is executed, where $1\leq i\leq |\mathscr R^*|$. We have 

\begin{lemma}[Subdivisions pruning]\label{lemma:prune unrelated subdivision}

Assume that $|o.\odot-r|>1$, all subdivisions except $d^e$ can be pruned safely.   
\end{lemma}
\noindent \textbf{Proof.}
By Lemma \ref{lemma:numbersubdivision1}, we only need to show the uncertainty region $u\subseteq d^e$. This can be proved based on Lemma   \ref{lemma:choose real ur} and  \textit{analyst geometric};    omitted due to space limit.
$\square$


Let $d_b^e$ be the MBR of the effective subdivision $d^e$. 
Lemma \ref{lemma:numbersubdivision1} and \ref{lemma:prune unrelated subdivision} indicate that we can immediately  discard unrelated subdivisions  once multiple subdivisions appear. In particular, we can  use   $d_b^e$ to prune the rest of candidate restricted areas, as it has a stronger pruning power compared to   $o.\odot_b$. We remark that the entity $o.\odot$ is  continuous evolving when it subtracts candidate restricted areas one by one, and in fact we  use $e$ (rather than $o.\odot$) to do ``subtraction operation'', as we adopt the \textit{pre-approximation} strategy.  We abuse the notation $o.\odot$ in Definition \ref{definition:effective subdivision} and Lemma \ref{lemma:numbersubdivision1} and \ref{lemma:prune unrelated subdivision}.  

\noindent \textbf{Comparison.} 
The approach above  is superior to the approach in Section \ref{subsec:pickout} (called the \textit{prior approach}) in the following points (note that the prior approach chooses the real uncertainty region at the last step): 

\noindent \textbf{1.} The prior approach needs to use \textit{each} subdivision to subtract  the rest of candidate restricted areas. In contrast, the approach above only needs to use  $d^e$ to subtract the rest of candidate restricted areas. For instance, in Figure \ref{fig:4b}, the prior approach  uses not only $d_1$ but also $d_2$ to subtract the rest of candidate restricted areas ($r_2$, $\cdots$, $r_7$), whereas the approach above only needs to use $d_1$ to subtract the rest  of candidate restricted areas.

\noindent \textbf{2.} The prior approach cannot prune the rest of candidate restricted areas. In contrast, the approach above can prune the unrelated candidate restricted areas. For example, in Figure \ref{fig:4b}, the prior approach cannot prune  candidate restricted areas as they are related to either $d_1$ or $d_2$, whereas the approach above can use the MBR of $d_1$ to prune $r_2$ and $r_6$, and use $d_1$ to prune $r_7$. 

\begin{figure}[t]
  \centering
  \subfigure[\scriptsize ]{\label{fig:4a}
     \includegraphics[scale=.45]{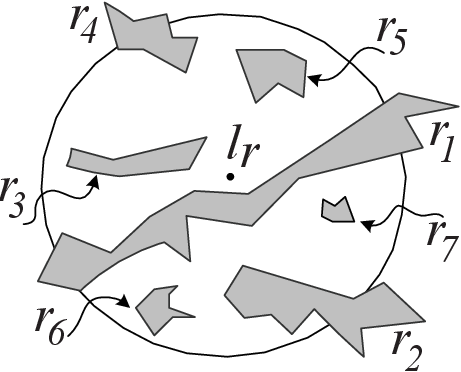}} 
     \hspace{3ex}
  \subfigure[\scriptsize ]{\label{fig:4b}
      \includegraphics[scale=.44]{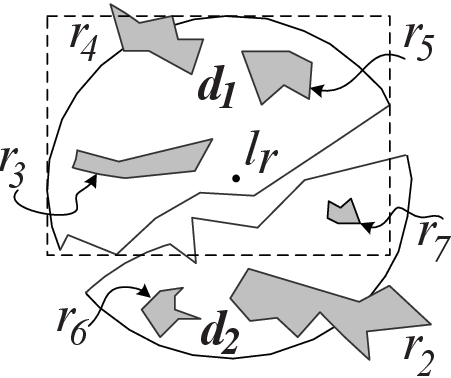}} 
 \caption{\small Illustration of subdivisions pruning  and span. The grey polygons illustrate the set $\mathscr R^*$ of candidate restricted areas, and the big equilateral polygon illustrates $e$ with 32 edges.} 
 \label{fig:4}
\end{figure}

We show the superiority of the approach above. The natural method  to compute $u$ however, is using $e$ to \textit{randomly} subtract each $r$ ($\in \mathscr R^*$)   one by one (cf. Section \ref{subsec:pickout}), implying that $r_1$ in Figure \ref{fig:4} may be  handled at last. In this case, the superiority of the approach above disappears. In the next section, we show how to  maximize its superiority by utilizing the span.

\subsection{Span} \label{subsec:span}

Let $g$ be a 2D  entity, and $g_b$ be the MBR of $g$. Let $(g_b.x^-,~g_b.y^-)$ and $(g_b.x^+,~g_b.y^+)$ be the left-bottom point  and right-top point of $g_b$, respectively.  We denote by $g_s$ the span of $g$, which is defined as follows.

\begin{definition}[Span]

Given a 2D entity $g$, its span $g_s$ is computed as 
\begin{subnumcases}{g_s=}
g_b.x^+ -g_b.x^-,&if $g_b.x^+ - g_b.x^-$ $\geq g_b.y^+ - g_b.y^-$ \nonumber  \\
g_b.y^+ -g_b.y^-,&otherwise \nonumber
\end{subnumcases}
 \end{definition}

\begin{heuristic}\label{heuristic:span}
A 2D entity with the larger span usually is more likely to subdivide a single (closed) region.
\end{heuristic}

See Figure \ref{fig:4a} for example.  Clearly, here $e$ can  be regarded as a single (closed) region, and each $r\in \mathscr R^*$ can  be regarded as a 2D entity. Compared to other candidate restricted areas, here $r_1$  has the largest span and it is more likely to subdivide $e$ into multiple subdivisions. Heuristic \ref{heuristic:span} motivates us to handle $r$ that has the larger span as early as possible. This can be achieved by sorting their spans according to the \textit{descending} order. We remark that  the span is a real number, hence the overhead to sort $|\mathscr R^*|$ candidate restricted areas is pretty small, and (almost) can be overlooked compared to the overhead to execute  $O(|\mathscr R^*|)$ times geometrical subtraction operations. 


\noindent \textbf{Another application.}
To compute $s$, the method in Section \ref{subsec:pickout} (called the \textit{prior method}) uses the intersection result, denoted by $\varphi$ , between $u_o$ and $R$ to subtract each hole $u_h$.  (Here $u_o$ refers to the outer ring of $u$.)  We now show how to use the span of hole to improve the prior method. Let $\mathscr H$ be the set of holes of $u$. 
\begin{lemma}[Subdivisions retaining]\label{lemma:retaining}
Given $\varphi$ and $u_h$, if $|\varphi-u_h|>1$, any subdivision of this subtraction result  cannot be discarded, where $|\cdot|$ denotes the number of subdivisions.
\end{lemma}
\noindent \textbf{proof.} 
By contradiction, assume the subdivision $d$ ($\in \varphi-u_h$)  can be discarded. This implies any point $p^\prime \in d$ can be discarded, i.e., $o$ cannot reach  $p^\prime$ (from $l_r$) if it does not walk out of $o.\odot$. (Note that $d$ possibly contains or intersects with other holes, but the case $d$ itself being  a hole is impossible. Otherwise, $d$ and $u_h$ form a larger hole as they are connected.) However, by the definition of $u$, $o$ can reach any point $p^\prime \in u_o-\bigcup _{u_h\in \mathscr H} u_h$ without the need of walking out of $o.\odot$,  it is contrary to the conclusion above. This completes the proof.  
$\square$

The lemma above indicates that once $|\varphi-u_h|>1$, we need to use each subdivision to subtract the rest of holes.   Without loss of generality, assume that  it produces $k$ subdivisions after we handle $|\mathscr H|-i$ holes, where $i\leq |\mathscr H|$.  For ease of discussion, we assume each hole $u_h$ at most can subdivide $\varphi$ into two subdivisions, and all the previous $|\mathscr H|-i$ holes can subdivide $\varphi$, implying that $k=|\mathscr H|-i+1$.  We can easily know that handling the previous $|\mathscr H|-i$ holes needs $1+2+\cdots+(k-1)$ $=\frac{(k-1)(k-2)}{2}$ times subtraction operations, since a new subdivision is to be produced when  a hole is handled. For the rest of holes, assume that each of them cannot subdivide $\varphi$, handling them needs $k\times i$ times subtraction operations. Let $x_1$ be the total number of the subtraction operations when handling all the $|\mathscr H|$ holes, we have  $x_1=\frac{(k-1)(k-2)}{2}+ki$. In contrast, if we swap the order to process the $|\mathscr H|$ holes. That is, we first handle  $i$ holes  that cannot subdivide $\varphi$ and then handle those $|\mathscr H|-i$ holes that can subdivide $\varphi$.  Similarly, let $x_2$ be the total  subtraction operation times. We have  $x_2=i+\frac{(k-1)(k-2)}{2}$.  Since $k=|\mathscr H|-i+1$, $x_1-x_2=|\mathscr H|i-i^2$. Hence, we have
\begin{equation}
\mathop{\mathrm{arg max} }_{i}  (x_1-x_2)=\frac{|\mathscr H|^2}{4}
\end{equation}
The formula above and Lemma \ref{lemma:retaining}  motivate us to handle holes that cannot subdivide $\varphi$ as early as possible. This can be achieved by  sorting their spans according to the \textit{ascending} order. For example, in Figure \ref{fig:7N} we handle $u_h^6$ and $u_h^7$ at last.  We remark that although some subtraction operations may be \textit{empty} operations when two entities are disjoint, it still incurs extra comparison overhead. In the sequel, we show two additional observations,  yielding two (small) mechanisms.

\begin{figure}[h]
\centering
\includegraphics[scale=0.6]{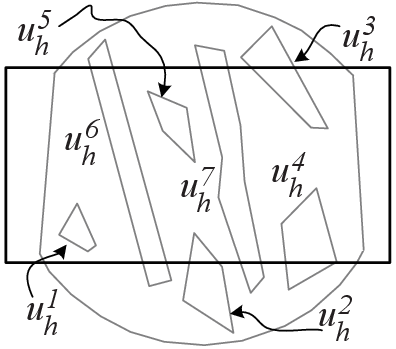}
\caption{\small Illustration of another application. The biggest rectangle denotes the query range $R$.  The uncertainty region $u$ is a closed region with 7 holes.}
\label{fig:7N}
\end{figure}

\noindent \textbf{Additional observations.} 
To compute  $u$, the method in Section \ref{subsec:pickout} is using $e$ to subtract each candidate restricted area $r$ on by one. Consider the case  $r\subset e$.   Clearly, $e-r$ forms a polygon with hole. For ease of discussion, let $e^\prime$ be the  subtraction result between $e$ and $r$, and assume that the next candidate restricted area to be processed is $r^\prime$. The natural approach is  using $e^\prime$ to subtract $r^\prime$. This approach however,  complicates the follow-up computation, and thus incurs extra overhead. This is mainly because geometrical operation on polygons with holes is generally more complicated and time consuming than on polygons without holes. To overcome this drawback, we employ a \textit{postpone processing} mechanism. Specifically, if $r\subset e$, we postpone the subtraction operation by caching $r$ in a temporary place; after all other candidate restricted areas are handled, we finally fetch $r$ from the temporary place and then handle it. For instance, in Figure \ref{fig:4b} we handle $r_3$ and $r_7$ at last.




Another common case is that  $r$ intersects with $e$ but $|e-r|=1$, where $|\cdot|$ denotes the number of subdivisions. To this case, the natural method is using $e$ to subtract $r$,  and then update the MBR of this subtraction result. This approach is inefficient, due to two main reasons: (\romannumeral 1)  such a new MBR usually  does not make enough contribution to the rest of computation, i.e., its pruning power is weak in most cases; (\romannumeral 2)  to obtain such a new MBR also needs to traverse the vertexes of this subtraction result, which incurs the extra overhead. To overcome this drawback, we employ a \textit{lazy update} mechanism. Specifically, if  $|e-r|=1$ (i.e., no multiple subdivisions appear),  we only execute the subtraction operation but do not  update the MBR of the subtraction result.  $r_4$ in Figure \ref{fig:4b}  illustrates this case, for example.  
We remark that the two mechanisms above can be directly applied to the case of computing $s$. For instance, see Figure \ref{fig:7N}, the \textit{lazy update} can be applied to $u_h^3$, and the \textit{postpone processing} can be applied to $u_h^1$.

\section{Precomputation based method}\label{sec:indexing}

In the previous discussion, we assume a twin-index is adopted:  $\mathscr I_r$ is used to manage  restricted areas, and $\mathscr I_o$ is used to manage    moving objects.  Once a moving object $o$ reports its new location to the server, we update  its recorded location $l_r$, and also update  $\mathscr I_o$. (See Section \ref{subsec:query processing}.) An obvious characteristic of this method    is  to compute  uncertainty regions {on the fly}, and an easily brought to mind method is to incorporate the precomputation strategy.

Simply  speaking,  we  index  restricted areas at first, and  then precompute  uncertainty regions and index them. Here we also adopt a twin-index. One is used to manage restricted areas, which is the same as the previous. Another, called $\mathscr I_{u}$, is used to manage the uncertainty regions.  
Specifically, for each moving object $o$,  we  search its candidate restricted areas on $\mathscr I_{r}$, and then compute its uncertainty region $u$ and  index it using $\mathscr I_{u}$.

Note that, it is possible that an object $o$   reports its new location to the server in the process of constructing $\mathscr I_{u}$. To this issue,   we differentiate two cases: (\romannumeral 1) the uncertainty region of this  object $o$ has ever been precomputed and indexed; and (\romannumeral 2) the uncertainty region of this object $o$ has not been precomputed.
Both of the cases can  be tackled easily.  For instance, regarding to the first case,  we  can update  its recorded location $l_r$ in the database, and then recompute its  uncertainty region  and update the current $\mathscr I_{u}$ right now. For the second case, we only need to update its recorded location $l_r$  in the database for the present. 
Once the precomputation is accomplished, the query can be executed, which is the similar as the previous. 
Henceforth, if an object $o$ reports its new location to the server, we also compute its uncertainty region  (off-line) and update  $\mathscr I_{u}$.  Note that although the precomputation based solution seems to be  more efficient, it however has a (non-trivial) drawback, i.e., its preprocessing time is rather large, which will be demonstrated in the next section.

\section{Performance study} \label{sec:6}

\subsection{Experiment settings}\label{subsec:experiment settings}
\noindent \textbf{Datasets}. 
In our experiments, both real and synthetic datasets are used. 
Two real datasets are named as CA and LB{\small \footnote{\small The CA dataset is available in site: \url{http://www.cs.utah.edu/~lifeifei/SpatialDataset.htm}, and the LB dataset is available in site: \url{http://www.rtreeportal.org/}. }}, respectively. 
The CA contains  104770  2D points, and the LB contains  53145  MBRs. We use the CA  to denote recorded locations of moving objects, and the LB  to denote  restricted areas. In order to simulate  moving objects with different characteristics, we randomly generate different distance thresholds (from 20 to 50) for them.  The size of 2D space is fixed at 10000$\times$10000, all datasets are normalized in order to fit this size of 2D space. Synthetic datasets  also include two types of information. We generate a number of polygons to denote  restricted areas, and let them uniformly distributed in  2D space.  We generate a number of points to denote recorded locations of moving objects, and let them randomly distributed in   2D space. Note that, there is an extra constraint --- these points cannot be located in any restricted area{\small \footnote{\small In fact, once this constraint is employed, the number of effective 2D points in the CA is 101871. Furthermore, since some MBRs in the LB are line segments, or they are not disjoint,  the number of effective rectangles in the LB is 12765 after we remove those unqualified MBRs.  }}.  We use the RE  and SY to denote the \underline{r}\underline{e}al and \underline{sy}nthetic datasets, respectively.

\noindent \textbf{Methods}. 
Existing methods    are invalid to our problem, we thus do not  compare  with them{\small \footnote{\small Imagine if we directly use existing methods (e.g.,  \cite{reynoldcheng:querying}), which renders the following unfair comparison: (\romannumeral 1)  the query answer  is clearly incorrect, as  analysed in Section \ref{sec:1}; and (\romannumeral 2)  the query time  is clearly less than  our algorithm's, as existing methods do not need to handle restricted areas.}}. The straightforward method  is infeasible and difficult to  implement, as  analysed in Section \ref{sec:frameworkandanalysis},  we thus do not discuss this method in our experiments (we believe the readers can understand this situation{\small \footnote{\small From another perspective,  the problem studied is different from most problems for which a straightforward, easy to implement and exact   method can always be found.}}).  Specifically, we implemented our \underline{s}olution (Section \ref{sec:our solution}), our \underline{s}olution together with the \underline{o}ptimization (Section \ref{sec:query evaluation}), and the \underline{p}recomputation based \underline{s}olution together with the \underline{o}ptimization (Section \ref{sec:indexing}). We use the same indexing structure, R-tree, for  the three algorithms above.  For brevity, we use {S, SO, and PSO} to denote them, respectively.  Furthermore, by the convention,  we implemented a \underline{b}aseline method that is to do a linear scan when searching candidate moving objects and candidate restricted areas (note: other strategies are the same as the ones of the S).  We use B to denote it for short.

\noindent \textbf{Distributions}. 
In our experiments, two types of  PDFs are used: uniform distribution and {distorted Gaussian} (note: our solutions can also work for other distribution PDFs, since we adopt the Monte Carlo method that can work for arbitrary distribution PDF). The definition of  distorted Gaussian is based on the  {general Gaussian}{\small \footnote{\small The general Gaussian has an infinite input space that is symmetric, its  PDF  is $\frac{1}{2\pi{\delta}^2}e^{\frac{(x-u_x)+(y-u_y)}{-2{\delta}^2}}$.  The input space of  {distorted  Gaussian}, however, is limited to the uncertainty region $u$ and it may be not symmetric.}}. Let  $pdf_G(x,y)$ and $pdf_{DG}(x,y)$   be the PDFs of general Gaussian and {distorted Gaussian}, respectively,  and let $\lambda$ be a coefficient, where  
$\lambda=\int_{\forall(x,y)\in u}pdf_G(x,y)dxdy$, we  have

\begin{subnumcases}{pdf_{DG}(x,y)=}
\frac{pdf_G(x,y)}{\lambda},&if $(x,y)\in u$\\
0,&otherwise
\end{subnumcases}

In theory, we should have calculated  $\lambda$ and converted  $pdf_{G}(x,y)$ into $pdf_{DG}(x,y)$ for each object $o$. Fortunately, we  need to neither calculate $\lambda$, nor  do any conversion. This is because   $\lambda$ will be eliminated when we substitute $pdf_{DG}(x,y)$ with $\frac{ pdf_{G}(x,y)}{\lambda}$ in the following formula.

\begin{equation}
p=\frac{\sum_{i=1}^{N_2}{pdf_{DG}{(x_i,y_i)}}}{\sum_{i=1}^{N_1}{pdf_{DG}{(x_i,y_i)}}}
\end{equation}
where $N_1$, $N_2$ are the number of random points being located in $u$ and  $s$, respectively. For brevity, we use  UD and DG to denote \underline{u}niform \underline{d}istribution and \underline{d}istorted \underline{G}aussian, respectively.

\noindent \textbf{Metrics}. 
The performance metrics in our experiments include:  the I/O time, query time (the sum of I/O and CPU time), preprocessing time and accuracy. 
We use the workload error to measure the accuracy. 
Two types of common workload errors are the {relative workload error} ({{RWE}}) and   {absolute workload error} ({{AWE}}){\small \footnote{\small  $RWE=\frac{\mid estimated~ value - real~ value \mid}{real~ value}$,  $AWE= $ $ \mid estimated ~value - real ~value \mid$.}}. 
In order to investigate  I/O and query time, we  randomly generate  50 query ranges, and  run 10 times for each  test, and finally compute the average I/O and query time for estimating a single query. We run 10 times and compute the average value for estimating the preprocessing time. 
In order to get the workload error, we  generate an object $o$   at the centre of  the $2$D space, and assign a value to the distance threshold $\tau$, and then compute its uncertainty region $u$. Next, we generate  100  query ranges that have the same size,  but have different intersections with  $u$. At first run, we get the real answer  of each query by setting  $N^\prime=1e+7$. (We remark that an \textit{absolute} real answer is unavailable, since the Monte Carlo method itself is an approximation algorithm. Even though, this obtained answer  can  be almost regarded as  the  real value, as we assign  a very large number to  $N^\prime$. In the rest of the paper, we slightly abuse the term “real value”.)  Next, we vary the size of $N^\prime$ to get several groups of workload errors.  We note that  another parameter $\xi$ also results in workload errors. When we study the impact of $\xi$ on the accuracy, we also use the workload error to estimate the returning results. The test method is similar to the one used to test $N^\prime$. Specifically, we get the real answer of each query by setting $\xi=32^2$, then vary $\xi$ to get several groups of workload errors.


\begin{table}[h]
\begin{center}
\begin{tabular}{|p{.02\textwidth} p{.29\textwidth} p{.30\textwidth} | }\hline 
{{\footnotesize Para. }}&{{\footnotesize Description}}& {{\footnotesize Value}}  \\ \hline 
{\footnotesize $\xi$}  & {\footnotesize  number of edges of $e$}		&{\footnotesize \textbf{[}$16,24, \textbf{32},48,64,32^{2}$\textbf{]}}  \\
{\footnotesize $N$}  & {\footnotesize cardinality of $\mathscr O$}			&{\footnotesize \textbf{[}$ 10k,20k,30k,40k,\textbf{50k}$\textbf{]}} \\
{\footnotesize $M$}  &{\footnotesize cardinality of $\mathscr R$}	&{\footnotesize \textbf{[}$10k,20k,30k,40k,\textbf{50k}$\textbf{]}} \\
{\footnotesize $\theta$}  &{\footnotesize size of $R$}	&{\footnotesize \textbf{[}$100,200,300,400,\textbf{500}$\textbf{]}} \\


 {\footnotesize $\zeta$}  &{\footnotesize number of edges of $r$}	&{\footnotesize \textbf{[}$\textbf{4},8,16,32,64$\textbf{]}}\\
{\footnotesize $N^{\prime}$}  &{\footnotesize number of {random points}}	&{\footnotesize \textbf{[}$600, \textbf{700},800,900,5k,6k,10^7$\textbf{]}}\\
 {\footnotesize PDF}  &{\footnotesize  distribution in $u$}	&{\footnotesize \textbf{[} \textbf{UD}, DG \textbf{]}}\\
 {\footnotesize $\tau$}   &{\footnotesize distance threshold of $o$} &{\footnotesize \textbf{[}$20,21,\cdots,49,50$\textbf{]}}\\
 \hline
\end{tabular}
\end{center}
\caption{\small Parameters Used in Our Experiments}\label{tab:experiment_parameters}
\end{table}

\noindent \textbf{Parameters}. 
All codes used in our experiments are written in C++ language; all experiments are conducted on a computer with 2.16GHz dual core CPU and 1.86GB of memory, running Windows XP. The page size is fixed to 4096 bytes; the maximum number of children nodes in the R-tree is fixed to 50.  The standard deviation of $pdf_G(x,y)$ (used for defining  $pdf_{DG}(x,y)$) is set to $\frac{\tau}{5}$, the mean $u_x$ and $u_y$ are set to $l_r.x$ and $l_r.y$, respectively, where ($l_r.x$, $l_r.y$) denote the coordinates of $l_r$.

The settings of other parameters  are illustrated in Table \ref{tab:experiment_parameters}, in which the numbers in \textbf{bold} represent the default settings.  $N$, $M$ and $\zeta$  refer to the  settings of  synthetic datasets, the default setting of each restricted area $r$ is a rectangle with $40\times 10$ size, and the one of the query range $R$ is a rectangle with $500\times 500$ size. 


\subsection{Results}\label{subsec:result}

\subsubsection{Choose  the error metric and the size of $N^\prime$}\label{subsec:reasonAbsoluterror}

Recall Section \ref{subsec:experiment settings} that there are two error metrics, both of them can be used to measure the accuracy. Note that, to ensure a  small  RWE  takes  more time (a non-trivial number)  than to ensure a same value of {{AWE}}. The results shown in Figure \ref{fig:exp:7a} confirm this fact.  These results are derived by setting  $\tau=20$. In this figure,   the AWE is $0.95\%$ (i.e., 0.0095) and the {RWE} is $10.75\%$, when $N^{\prime}=700$.  It is unreasonable if we choose $10.75\%$ as the {RWE}. Otherwise, it implies  that  returning a value of $89.25\%$ will be tolerated even if the real value is $100\%$. Therefore,  we need to increase $N^{\prime}$ in order to get a smaller {RWE}. By doing so, we get {{RWE}}$=1.12\%$  and {{AWE}}$=0.05\%$ at $N^{\prime}=50000$. Therefore, if we want to assure a value $1\%$ of RWE, we  have to set $N^{\prime}>50000$ at least. However, even if we let $N^{\prime}=50000$, and only  compute a single object's probability, it  takes about 2871 milliseconds, implying that to assure a small RWE (e.g., $1\%$) takes much time. To further verify this fact, we conduct another set of experiments using both real and synthetic datasets where we set $N^\prime=50000$  and others are the default settings. The results are listed in the  table below.

\begin{table}[h]
\begin{center}
\begin{tabular}{p{.08\textwidth} p{.15\textwidth} p{.15\textwidth}  }\hline 
{{\footnotesize Dataset. }}&{{\footnotesize Total test time (sec.)}}& {{\footnotesize Query time (sec.)}}  \\ \hline



{\footnotesize RE}  &{\footnotesize 21399.25 }	&{\footnotesize  427.985 }\\
 {\footnotesize SY}   &{\footnotesize 166022.6} &{\footnotesize 332.004}\\
 \hline
\end{tabular}
\end{center}
\end{table}

The total test time   takes  about 59 and 46 \textit{hours}, respectively; and each single query takes several \textit{minutes}.  In view of these, and in most cases a small {AWE}  is enough to  satisfy our demand, in the rest of experiments we choose to assure a small  {AWE} by setting $N^\prime$ to a smaller value. Figure \ref{fig:exp:7b} depicts the results by setting   $\tau$ to 20, 30, 40 and 50, respectively. We can see  that on the whole, an object with a smaller $\tau$ usually needs a larger $N^{\prime}$, if we want to assure the same value of {AWE}. Based on the facts above,  unless stated otherwise,  we choose $N^{\prime}=700$ in the rest of experiments, which can ensure a value 0.01 of {{AWE}}. 

\begin{figure}[t]
\centering
  \subfigure[\scriptsize {AWE and RWE vs. $N^{\prime}$}]{\label{fig:exp:7a}
     \includegraphics[width=23.1ex,height=17.4ex]{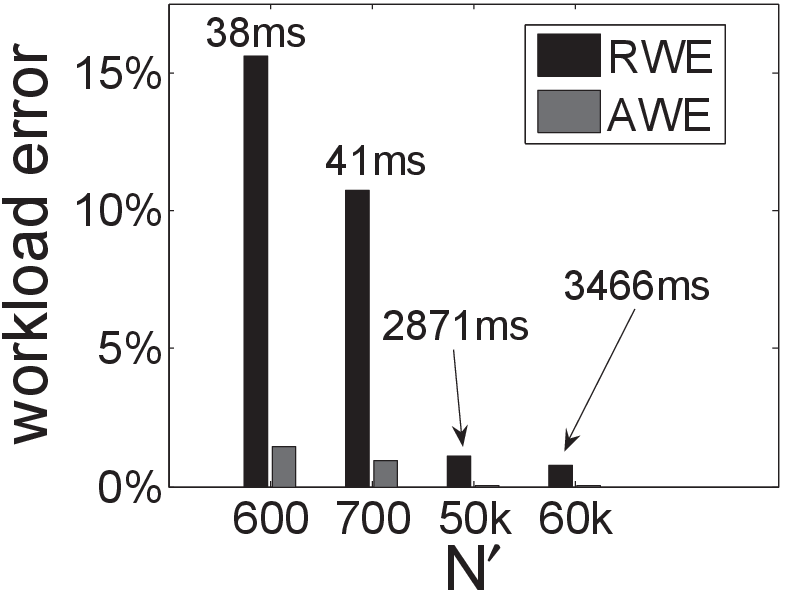}} 
  \subfigure[\scriptsize {AWE vs. $\tau$ and $N^{\prime}$}]{\label{fig:exp:7b}
     \includegraphics[width=24.1ex,height=17.4ex]{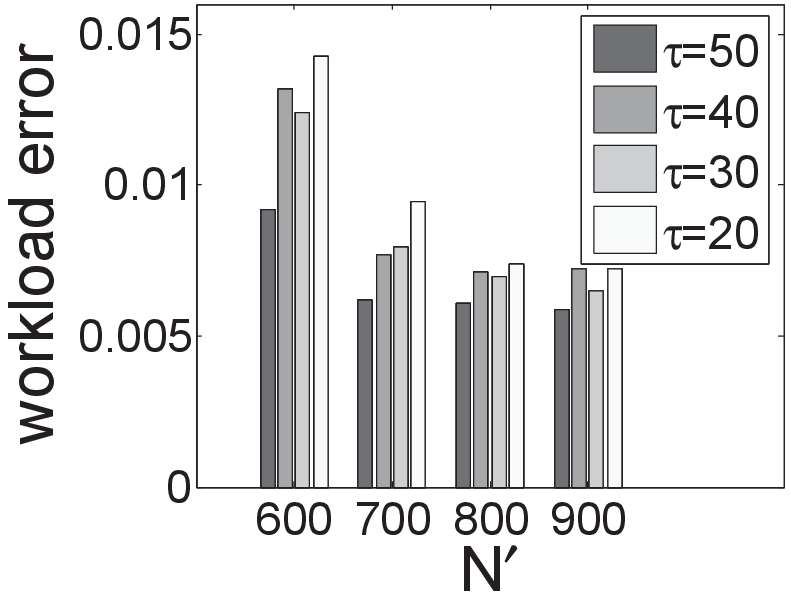}} 
\caption{\small Workload Error Comparison.}
\label{fig:exp:7}
\end{figure}

\begin{figure}[t]
\centering
   \subfigure[\scriptsize {SY dataset (query time) }]{\label{fig:exp:5a}
      \includegraphics[width=23.1ex,height=17.4ex]{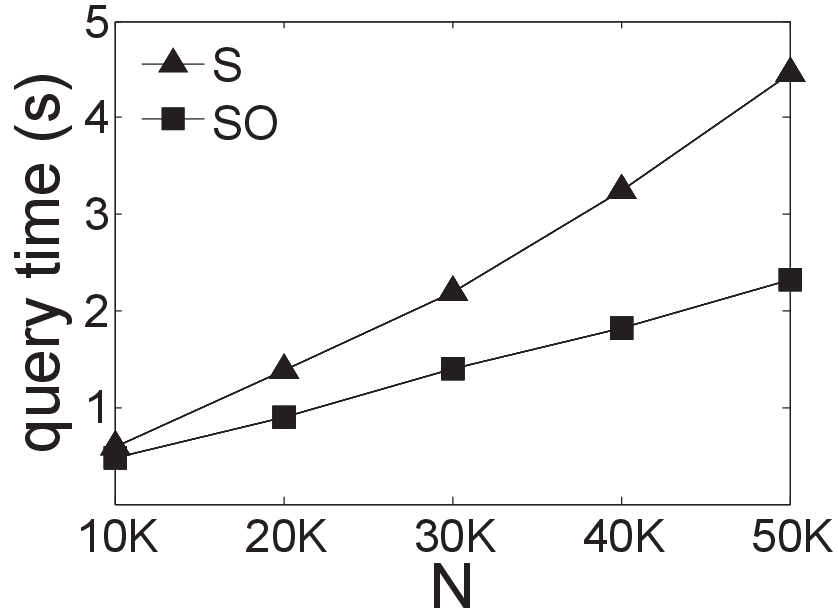}}
  \subfigure[\scriptsize {SY dataset (I/O time)} ]{\label{fig:exp:5b}
     \includegraphics[width=23.1ex,height=17.4ex]{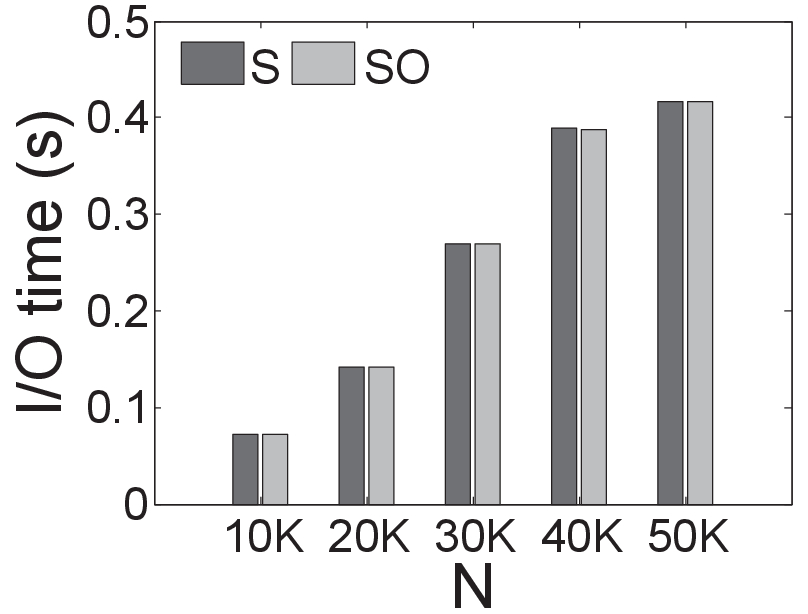}}
\caption{\small Query  and I/O performance vs. $N$.}
\label{fig:exp:5}
\end{figure}

\subsubsection{S vs. SO} \label{subsubsub:two online}

We first study the impact of $N$, $M$ and $\zeta$ based on synthetic datasets, and then study the impact of $\theta$ and $\xi$ based on both real and synthetic datasets.


\noindent  \textbf{Impact of $N$}.
Figure \ref{fig:exp:5} illustrates the results by varying $N$ from 1e$+4$ to 5e$+4$. We can see that with the increase of $N$, both the query and I/O time increase for the two methods.  In terms of query performance, we find that the SO outperforms the S, which  demonstrates the efficiency of our optimization. 
In particular, their performance differences   are more obvious especially when $N$ is large, which demonstrates  the scalability of SO  is better than the one of  S.  We remark that  the I/O performance of the two methods is identical, this is because same location  and restricted area records are fetched from the database for the two  methods. In the rest of experiments, we only report the IO performance of  SO.






\begin{figure*}[t]
  \centering
  \subfigure[\scriptsize { SY dataset (query time)} ]{\label{fig:exp:111a}
    \includegraphics[width=23.1ex,height=17.4ex]{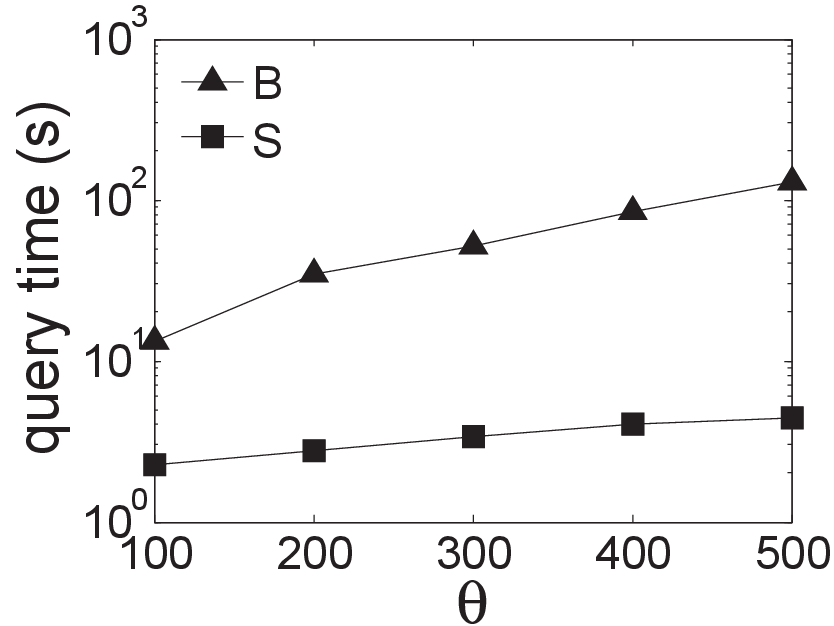}}  \hspace{1ex}
  \subfigure[\scriptsize {SY dataset (I/O time)} ]{\label{fig:exp:111b}
      \includegraphics[width=23.1ex,height=17.4ex]{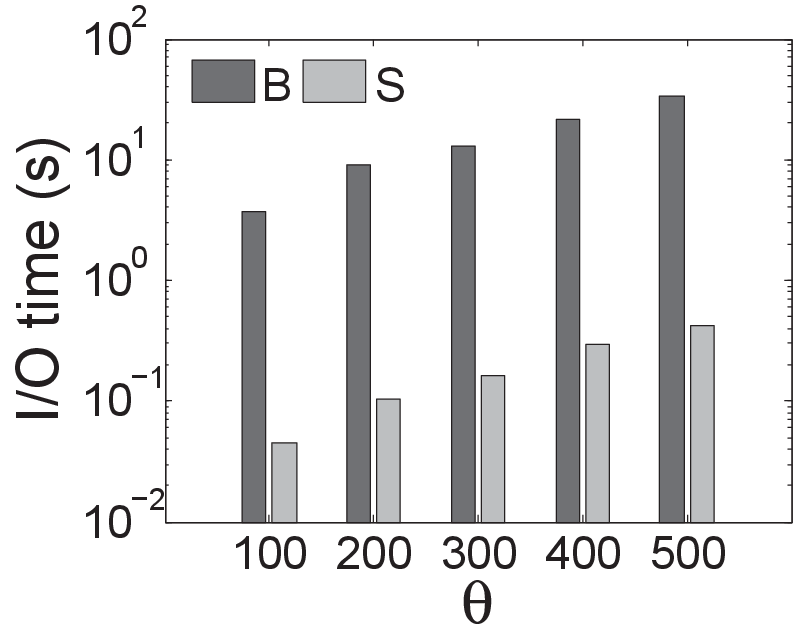}}  \hspace{1ex}
  \subfigure[\scriptsize {RE dataset (query time)} ]{\label{fig:exp:111c}
      \includegraphics[width=23.1ex,height=17.4ex]{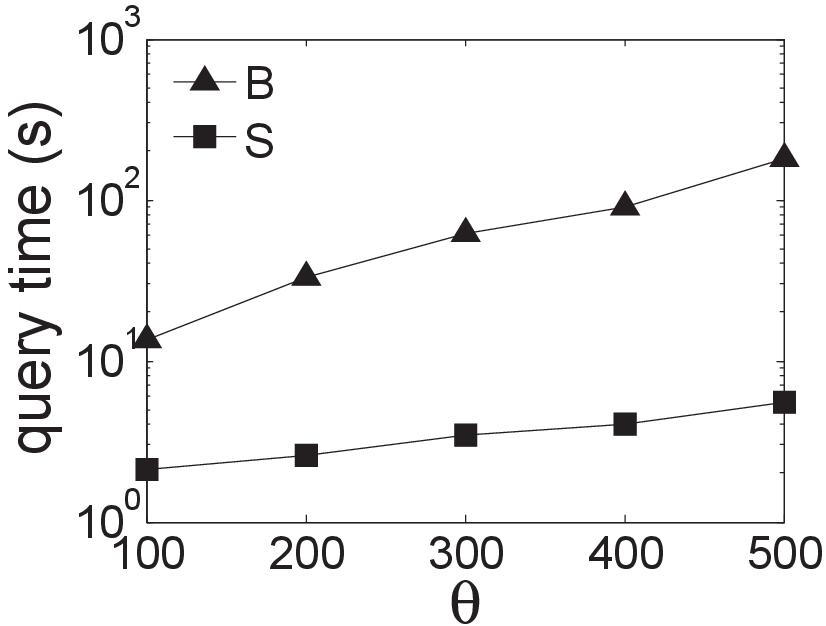}} \hspace{1ex}
  \subfigure[\scriptsize {RE dataset (I/O time)} ]{\label{fig:exp:111d}
      \includegraphics[width=23.1ex,height=17.4ex]{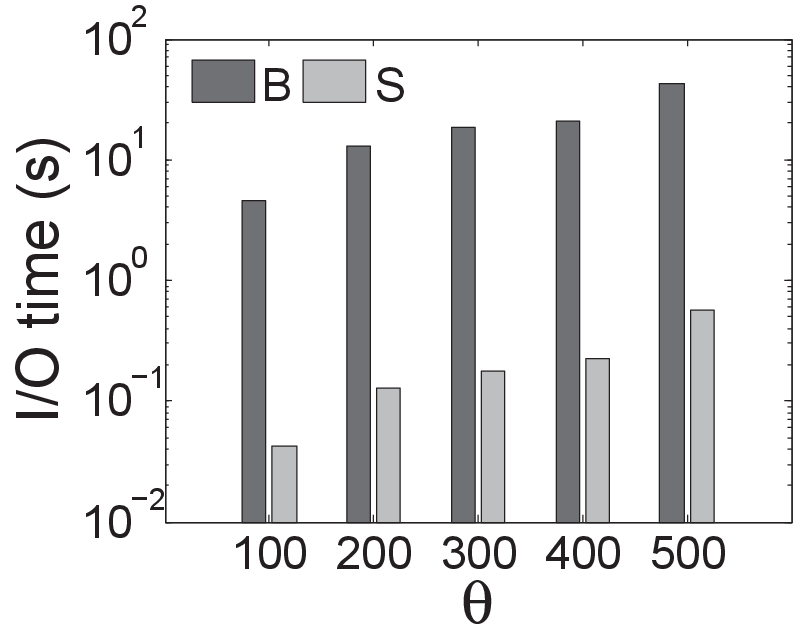}}
 \caption{\small Query and I/O performance vs. $\theta$.} 
 \label{fig:exp:11}
\end{figure*}

\begin{figure*}[t]
  \centering
  \subfigure[\scriptsize {SY dataset (query time)} ]{\label{fig:exp:9a}
    \includegraphics[width=23.1ex,height=17.4ex]{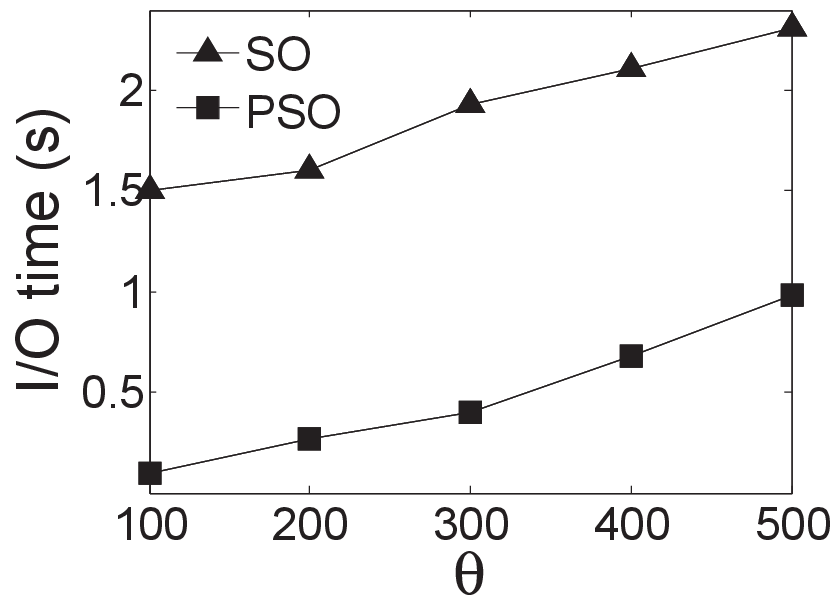}} \hspace{1ex}
  \subfigure[\scriptsize {SY dataset (I/O time)} ]{\label{fig:exp:9c}
      \includegraphics[width=23.1ex,height=17.4ex]{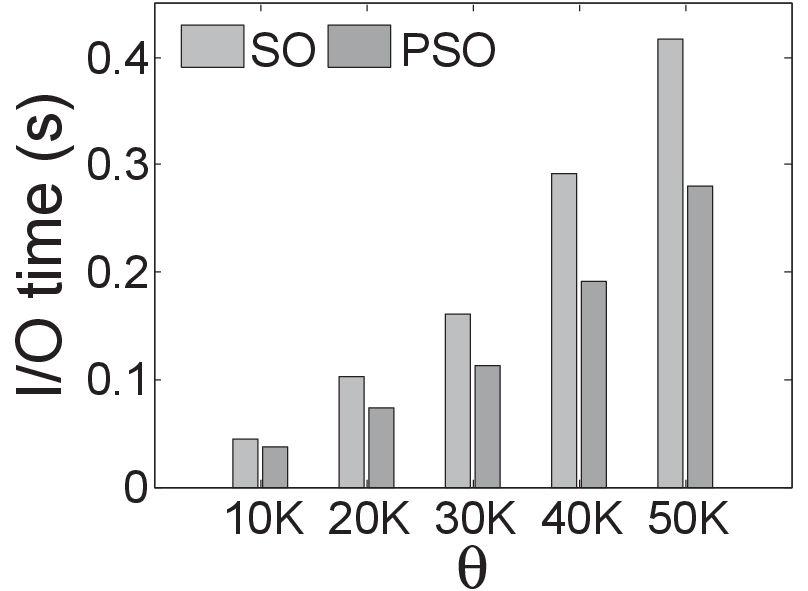}} \hspace{1ex}
  \subfigure[\scriptsize {RE dataset (query time)} ]{\label{fig:exp:9d}
      \includegraphics[width=23.1ex,height=17.4ex]{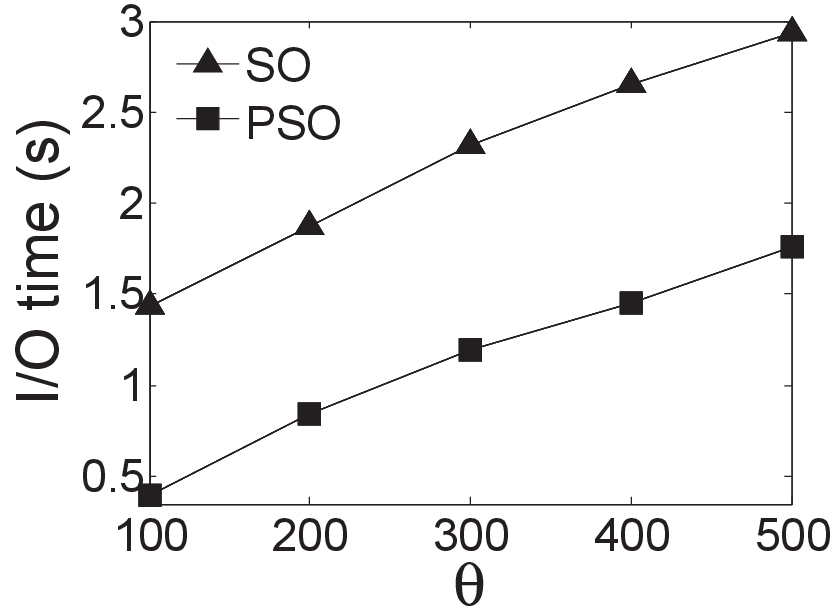}} \hspace{1ex}
  \subfigure[\scriptsize {RE dataset (I/O time)} ]{\label{fig:exp:9f}
      \includegraphics[width=23.1ex,height=17.4ex]{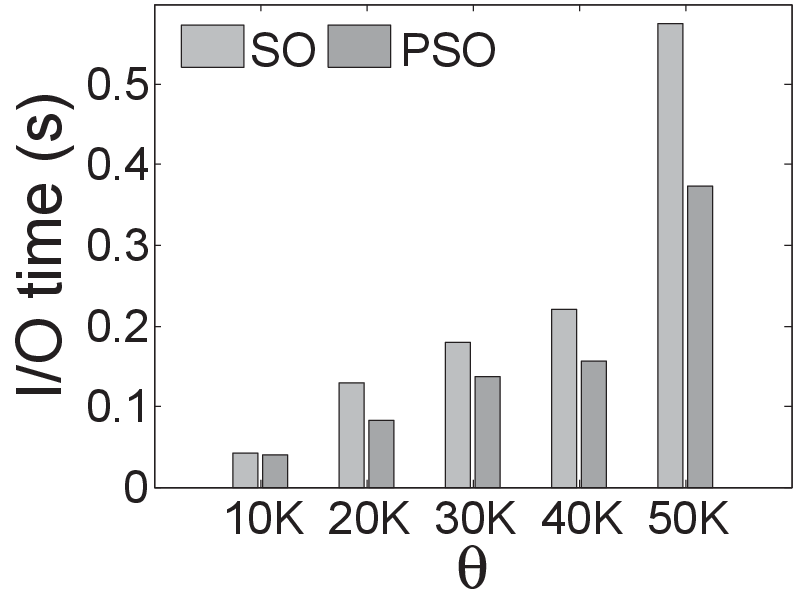}}
 \caption{\small Query and I/O performance vs. $\theta$.} 
 \label{fig:exp:9}
\end{figure*}


\subsubsection{B vs. S} \label{subsubsub:baseline}
We can see from Figure \ref{fig:exp:11} that both the query and I/O performance of the S significantly  outperform the ones of B. Note that when we vary other parameters in addition to $\theta$, we also get  similar results, i.e., the S significantly outperforms the B. Clearly, the SO also significantly outperforms the B as it is superior than the S.


\subsubsection{SO vs. PSO}\label{subsubsec:preprocess and on}
From Figure \ref{fig:exp:9}, we can  see that the PSO outperforms the SO regardless of query or I/O performance, which demonstrates the benefits of precomputing uncertainty regions. Note that we also vary other parameters and find that the PSO also outperforms the SO in terms of query and I/O performance.

However, we find that  the  time for precomputing uncertainty regions is  rather long, the results are plotted in Figure \ref{fig:exp:10}.  The PSO takes 2532.828 seconds (about 42 minutes) when the default settings of the synthetic datasets  are used (note: the SO do not need to precompute uncertainty regions, and  only need  to index restricted areas and moving objects, which can be finished in several seconds). In addition, when we set  $\zeta$ to 64,  the PSO takes 5386.812 seconds (about an hour and a half). The long preprocessing time can be regarded as a (non-trivial) drawback of this approach.  

\begin{figure}[b]
  \centering
   \subfigure[\scriptsize {SY dataset} ]{\label{fig:exp:10a}
      \includegraphics[width=16.2ex,height=16.2ex]{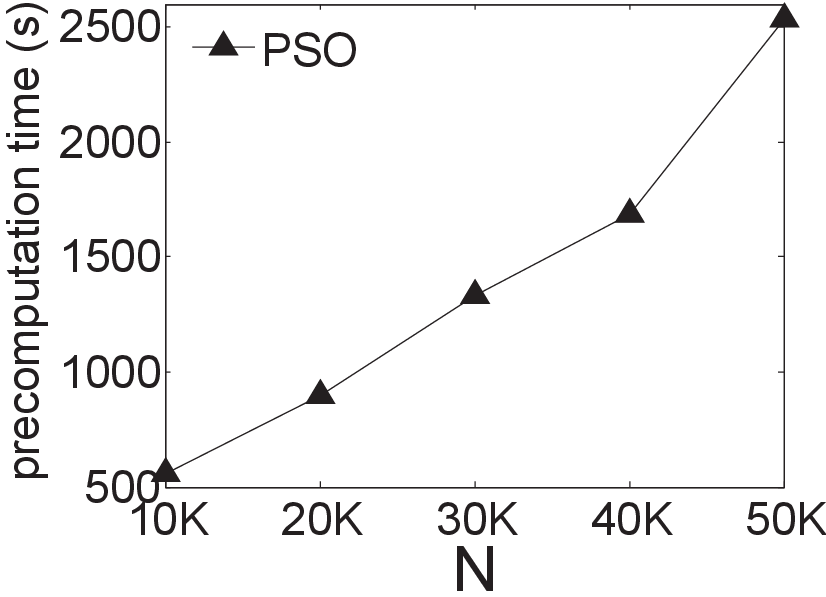}} 
   \subfigure[\scriptsize {SY dataset} ]{\label{fig:exp:10b}
       \includegraphics[width=16.2ex,height=16.5ex]{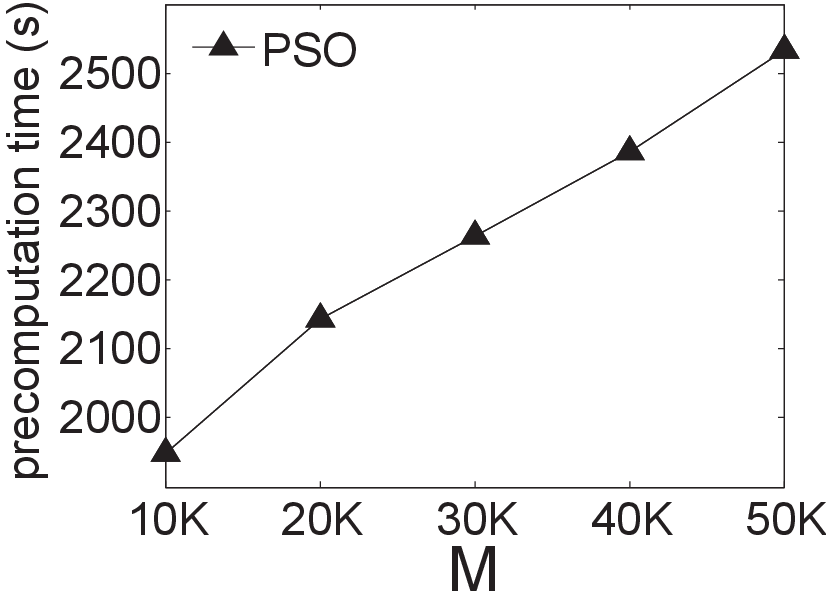}} 
   \subfigure[\scriptsize {SY dataset} ]{\label{fig:exp:10C}
      \includegraphics[width=16.2ex,height=16.2ex]{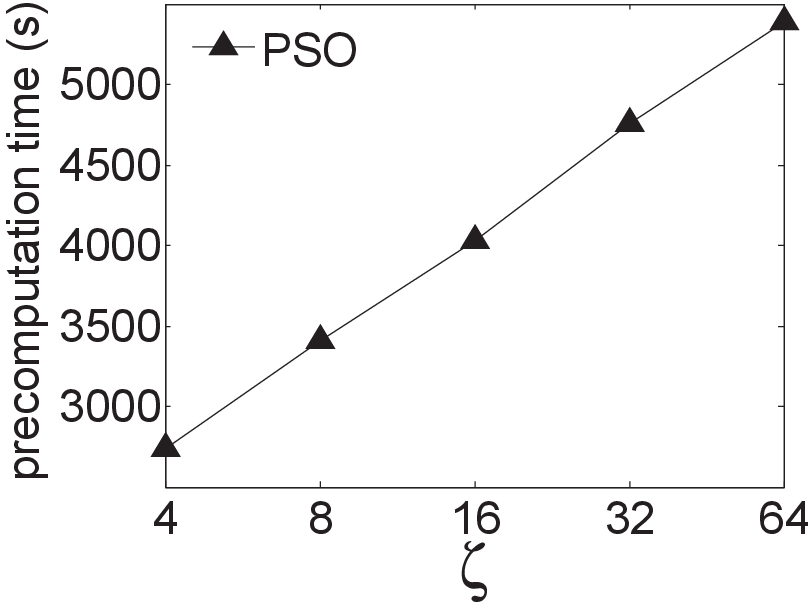}} 
 \caption{\small Precomputation time vs. $N$, $M$ and  $\zeta$. } 
 \label{fig:exp:10}
\end{figure}


\subsubsection{Compare  different  PDFs} 
We next  test the impact of PDFs. Specifically, we let all parameters be totally same except the PDF (note: here we use our preferred method, i.e., the SO).    On one hand, we compare their query time by  varying $\theta$ (here  $N^{\prime}$ is the default setting). On the other hand, we compare their accuracies  by varying $\xi$ (here we set $N^{\prime}$ to $10^{7}$).

\noindent \textbf{Query time.} 
Figure \ref{fig:exp:8a} and \ref{fig:exp:8b} depict the results when we vary $\theta$. We can see that the query time when the PDF is DG is  more than the one when the PDF is UD. This is mainly because the time  computing a single object's probability is relatively long when the PDF is DG.

\noindent \textbf{Accuracy.} 
In addition, by varying $\xi$ from 16 to 64, their accuracies are plotted in Figure \ref{fig:exp:8c}. As we expected, the larger (the) $\xi$ is, the more accurate  answer we can get. In particular, we can see that, compared to the case of uniform distribution, $\xi$ makes less impact on  the accuracy when the PDF is DG. Hearteningly, even if the PDF is UD, the accuracy of the proposed method is still high since the AWE is about 634$\times 10^{-6}$ when $\xi=32$.   Moreover, we can see from Figure \ref{fig:exp:8d} that,  with the same $\xi$,  the smaller the distance threshold $\tau$ is, the more accurate answer we can get when the PDF is UD. Interestingly, the case of DG is exactly the opposite, which confirms (in  a different way)  the previous conclusion derived from  Figure \ref{fig:exp:7b}.
Furthermore, we also report the   RWEs of the set  of experiments, which are shown in Figure \ref{fig:exp:8e} and \ref{fig:exp:8f}, respectively. By comparing Figure \ref{fig:exp:8c}/\ref{fig:exp:8d} and Figure \ref{fig:exp:8e}/\ref{fig:exp:8f}, we can easily see that in the same settings, the AWE is significant smaller than the RWE, which is in line with the conclusion derived from  Figure \ref{fig:exp:7a}.  
\begin{figure}[b]
  \centering
  \subfigure[\scriptsize SY dataset (query time)]{\label{fig:exp:8a}
      \includegraphics[width=22ex,height=16.5ex]{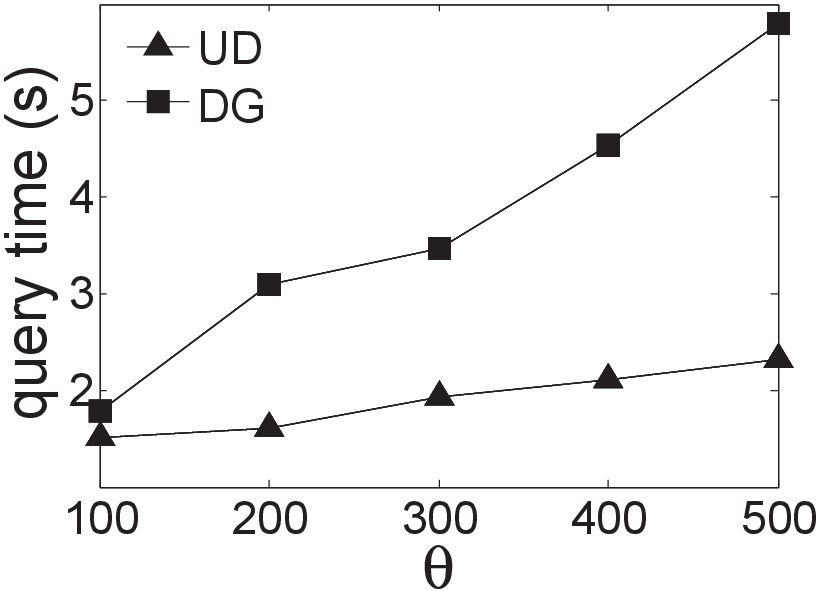}} \hspace{1ex} 
      \hspace{1ex}
   \subfigure[\scriptsize RE dataset (query time)]{\label{fig:exp:8b}
       \includegraphics[width=22ex,height=16.5ex]{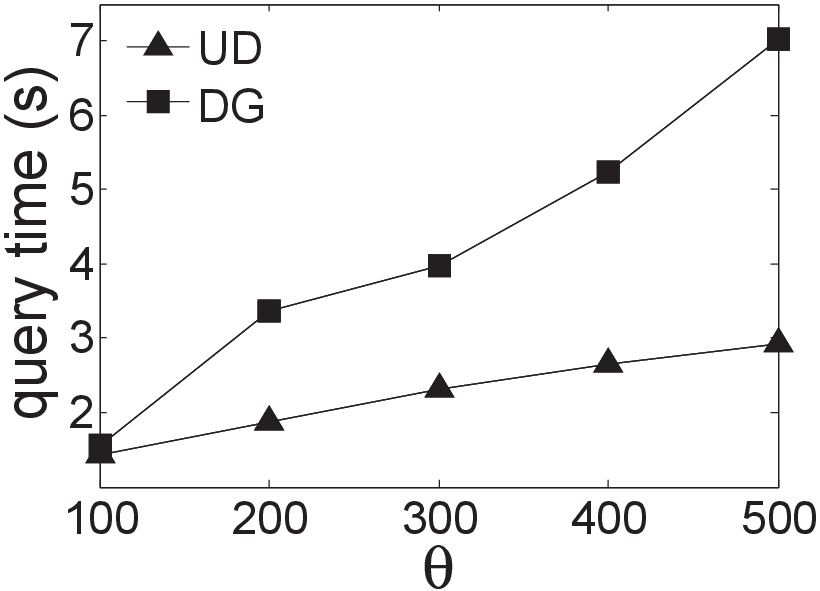}}  \hspace{1ex} 
 \caption{\small Query efficiency  vs. PDF. } 
 \label{fig:exp:8new}
\end{figure}

\begin{figure*}[t]
  \centering
  \subfigure[\scriptsize AWE vs. $\xi$ ($\tau=50$) ]{\label{fig:exp:8c}
      \includegraphics[width=23.1ex,height=17.4ex]{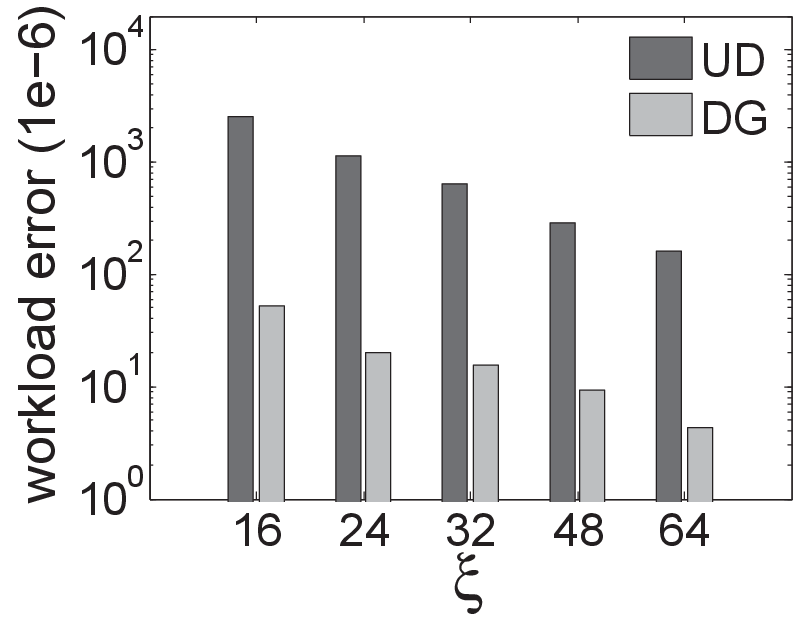}} \hspace{1ex}
   \subfigure[\scriptsize AWE vs. $\tau$ and $\xi$]{\label{fig:exp:8d}
       \includegraphics[width=23.1ex,height=17.4ex]{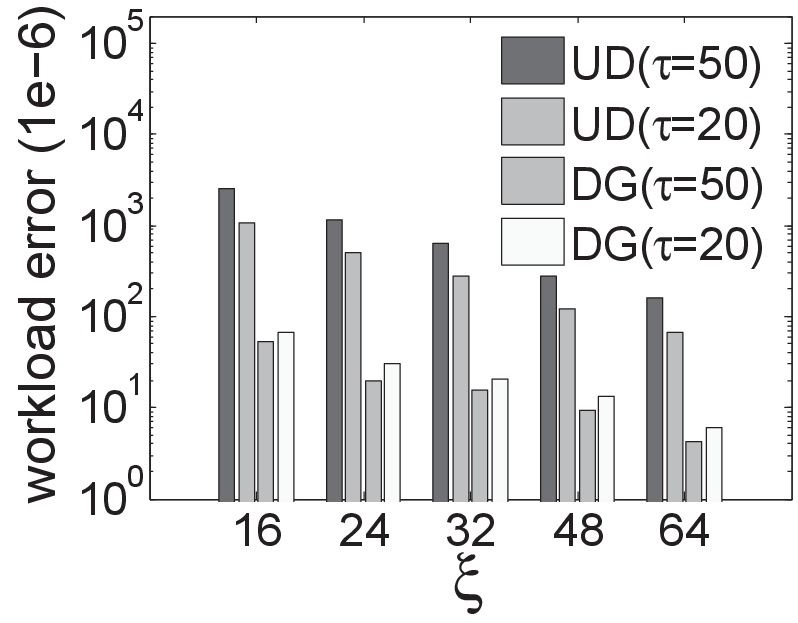}}
  \subfigure[\scriptsize RWE vs. $\xi$ ($\tau=50$)]{\label{fig:exp:8e}
      \includegraphics[width=23.1ex,height=17.4ex]{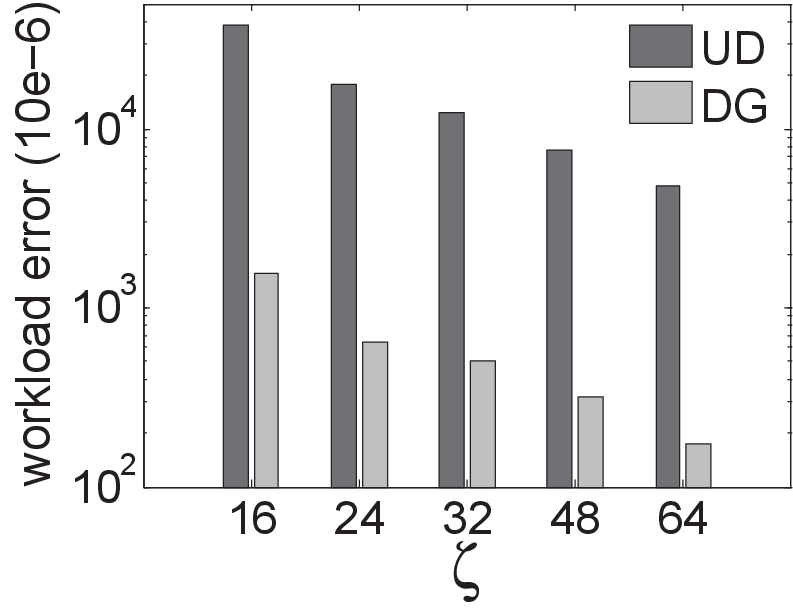}} \hspace{1ex}
   \subfigure[\scriptsize RWE vs. $\tau$ and $\xi$]{\label{fig:exp:8f}
       \includegraphics[width=23.1ex,height=17.4ex]{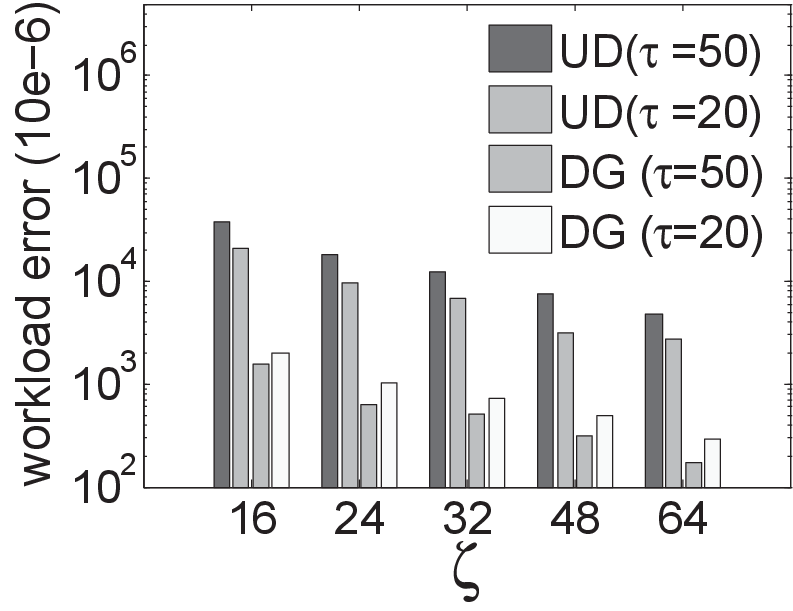}} 
 \caption{\small Accuracy vs. PDF.} 
 \label{fig:exp:8}
\end{figure*}

\section{Conclusion} \label{sec:7}

This paper studies the CSPRQ for uncertain moving objects. The deliberate analyses offer  insights into the  problem considered, and show  that to process the CSPRQ  using a straightforward method is infeasible.  We propose the targeted solution and  demonstrate  its  efficiency and  effectiveness  through extensive experiments. An additional finding is the precomputation based method has a non-trivial  preprocessing time (although it outperforms our preferred solution in other aspects),  which offers an important indication  sign  for the future research. We conclude this paper with several interesting  research topics: (\romannumeral 1) how to process the CSPRQ in  3D space?  (\romannumeral 2)  if the location update policy is the time based update, rendering that the uncertainty region $u$ is to be a continuously changing geometry over time, how to process the CSPRQ in such a scenario? (\romannumeral 3) if the query issuer is also moving,  the location of query issuer is also uncertain, how to process the location based CSPRQ?

\ifCLASSOPTIONcaptionsoff
  \newpage
\fi

{
\bibliographystyle{abbrv}
\bibliography{sample}
}


\end{document}